\documentclass[conference]{IEEEtran}
\IEEEoverridecommandlockouts
\usepackage{comment}
\usepackage[utf8x]{inputenc}
\usepackage{amsmath,amssymb,amsfonts}
\usepackage{amsthm} 
\theoremstyle{definition} 
\newtheorem{definition}{Definition}[section] 
\usepackage{algorithmic}
\usepackage{graphicx}
\usepackage{textcomp}
\usepackage{xcolor}
\usepackage{todonotes}
\usepackage{paralist}
\usepackage{numprint}
\usepackage{xspace}
\usepackage{cleveref}
\usepackage{multirow}
\usepackage{subcaption}
\usepackage{soul}
\usepackage{balance}
\usepackage[acronym,toc,shortcuts]{glossaries}
\usepackage{listings}
\usepackage{float}
\restylefloat{table}
\usepackage{fancyhdr} 
\usepackage[linesnumbered,ruled,vlined,lined,boxed]{algorithm2e}

\def\BibTeX{{\rm B\kern-.05em{\sc i\kern-.025em b}\kern-.08em
    T\kern-.1667em\lower.7ex\hbox{E}\kern-.125emX}}

\makeatletter
\newcommand\footnoteref[1]{\protected@xdef\@thefnmark{\ref{#1}}\@footnotemark}
\makeatother
\setdefaultenum{(i)}{}{}{}
\SetKwProg{Function}{function}{}{end function}
\SetKwInOut{KwVar}{Variables}
\SetKwInOut{KwDescribe}{Description}
\SetKw{KwTo}{to}
\SetKw{KwAnd}{and}
\SetKwIF{If}{ElseIf}{Else}{if}{then}{else if}{else}{end if}
\SetKwFor{For}{for}{do}{end for}
\SetKwFor{ForEach}{for each}{do}{end for}
\SetKwFor{ForAll}{for all}{do}{end for}
\SetKwFor{ForAllP}{for all}{do in parallel}{end for}

\def\BibTeX{{\rm B\kern-.05em{\sc i\kern-.025em b}\kern-.08em
    T\kern-.1667em\lower.7ex\hbox{E}\kern-.125emX}}

\newcommand{\mi}[1]{\mathit{#1}}

\newcommand{\TODO}[1]{\textbf{\color{red}\noindent (TODO: {#1})}\xspace}

\newcommand{\system}{\textsc{INetCEP}\xspace}
\newcommand{\Data}{\texttt{Data}\xspace}
\newcommand{\Interest}{\texttt{Interest}\xspace}
\newcommand{\DataStream}{\texttt{Data Stream}\xspace}

\newcommand{\AddQInterest}{\texttt{Add Query Interest}\xspace}
\newcommand{\RmQInterest}{\texttt{Remove Query Interest}\xspace}
\newcommand{\degree}{\ensuremath{^\circ}}
\newcommand{\eg}{e.g., }
\let\OldS\S
\renewcommand{\S}{\OldS\xspace}
\Crefformat{figure}{#2Fig.~#1#3}
\Crefmultiformat{figure}{Figs.~#2#1#3}{ and~#2#1#3}{, #2#1#3}{ and~#2#1#3}

\newcommand*\circled[1]{\tikz[baseline=(char.base)]{
            \node[shape=circle,draw,inner sep=0.3pt,color=white,fill=black] (char)
            {#1};}}

\makeglossaries
\newacronym{CEP}{CEP}{Complex Event Processing}
\newacronym{NFN}{NFN}{Named Function Networking}
\newacronym{IoT}{IoT}{Internet-of-Things}
\newacronym{NDN}{NDN}{Named Data Networking}
\newacronym{ICN}{ICN}{Information-Centric Networking}
\newacronym{CCN}{CCN}{Content-Centric Networking}
\newacronym{PIT}{PIT}{Pending Interest Table}
\newacronym{FIB}{FIB}{Forwarding Information Base}
\newacronym{CS}{CS}{Content Store}
\newacronym{OP}{OP}{Operator Placement}
\newacronym{QoS}{QoS}{Quality of Service}
\newacronym{INP}{INP}{in-network processing}

\begin{document}

\title{INetCEP: In-Network Complex Event Processing for Information-Centric Networking \\
\emph{(Extended Version)}
}

\author{\IEEEauthorblockN{Manisha Luthra\IEEEauthorrefmark{1},
Boris Koldehofe\IEEEauthorrefmark{1},
Jonas Höchst\IEEEauthorrefmark{2},
Patrick Lampe\IEEEauthorrefmark{2},
Ali Haider Rizvi\IEEEauthorrefmark{1},
Ralf Kundel\IEEEauthorrefmark{1},
Bernd Freisleben\IEEEauthorrefmark{2}
}
\IEEEauthorblockA{
\IEEEauthorrefmark{1}\textit{Technical University of Darmstadt, Germany} \\
\{firstname.lastname\}@kom.tu-darmstadt.de
}
\IEEEauthorblockA{
\IEEEauthorrefmark{2}\textit{University of Marburg, Germany} \\
\{lastname\}@informatik.uni-marburg.de
}
}

\maketitle
\thispagestyle{fancy}

\begin{abstract}
Emerging network architectures like \ac{ICN} offer simplicity in the data plane by addressing \emph{named data}. 
Such flexibility opens up the possibility to move data processing inside network elements for high-performance computation, known as \emph{in-network processing}.
However, existing \ac{ICN} architectures are limited in terms 
of (i) in-network processing and (ii) data plane programming abstractions. Such architectures can benefit from \ac{CEP}, an in-network processing paradigm to efficiently process data inside the data plane.
Yet, it is extremely challenging to integrate \ac{CEP} because the current communication model of \ac{ICN} is limited to \emph{consumer-initiated} interaction that comes with significant overhead in number of requests to process continuous data streams.
In contrast, a change to \emph{producer-initiated} interaction, as favored by \ac{CEP}, imposes severe limitations for request-reply interactions.

In this paper, we propose an in-network \ac{CEP} architecture, \system that supports unified interaction patterns (\emph{consumer-} and \emph{producer-}initiated).
In addition, we provide a CEP query language and facilitate \ac{CEP} operations while increasing the range of applications that can be supported by \ac{ICN}.
We provide an open source implementation and evaluation of \system over an \ac{ICN} architecture, Named Function Networking, and two applications: energy forecasting in smart homes and a disaster scenario.

\end{abstract}


\setlength{\textfloatsep}{0pt}
\section{Introduction} \label{sec:intro}

Emerging network architectures like Information-Centric Networking (\ac{ICN}) simplify the data plane of the current Internet by changing its addressing scheme from \emph{named hosts} to \emph{named data}.
\ac{ICN} has evolved as a key paradigm towards a content-centric Internet, as currently adopted by academia and industry, \eg by Internet2, Cisco, and  Intel~\cite{introCCN}
for real-world deployment~\cite{ICNDeploy2019}.
The data plane abstractions of \ac{ICN} are particularly useful since users can define \emph{what data they need} instead of identifying \emph{where to get it from}.
Additionally, exploiting data plane programmability on in-network elements of \ac{ICN} can offer high throughput by processing packets at line rate, while delivering them at low latency, typically known as \emph{in-network processing}. However, existing \ac{ICN} architectures like \ac{NDN} and \ac{NFN} are restricted in terms of data plane programmability due to lack of (i) in-network processing and (ii) data plane programming abstractions.

\begin{figure}[t]
	\centering
	\includegraphics[height=5cm]{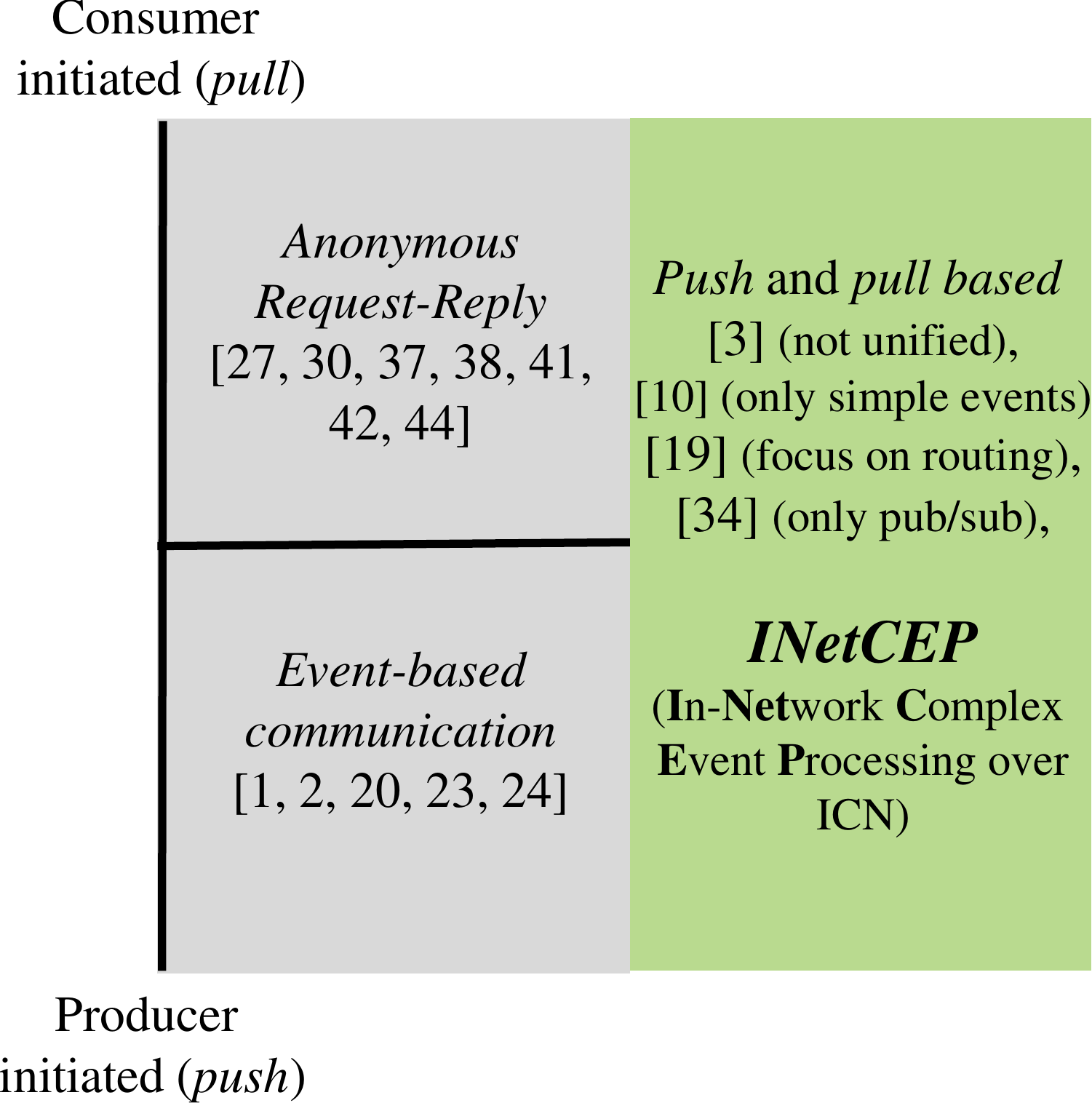}
	\setlength{\abovecaptionskip}{0pt}
	\setlength{\belowcaptionskip}{5pt}
	\caption{Taxonomy of ICN architectures based on the supported interaction patterns.}
	\label{fig:taxonomy}
\end{figure}

This makes Complex Event Processing (\ac{CEP}) a paradigm of choice for an \ac{ICN} architecture.
\ac{CEP} is a powerful \emph{in-network processing} paradigm that takes a \emph{query} as an input to describe the correlations over a set of incoming data streams in order to deliver data notifications in response to the query.
For instance, in a disaster scenario, a heat map \emph{query} can describe correlations over a set of data streams, \eg location updates from victims to deliver a heat map distribution of survivors to better coordinate the activities of rescue workers.

However, employing \ac{CEP} on top of an \ac{ICN} architecture is extremely challenging.
An important challenge is that the communication model of current \ac{ICN} architectures has strong limitations in supporting the processing of periodic data streams.
For instance, \ac{NDN} uses a \emph{consumer-initiated} interaction pattern where a consumer \emph{pulls} data by sending an \Interest (request) to the network.
The \ac{NDN} network forwards this request to one or more producers that satisfy the request and then forward the \Data (reply) back to the consumer.
For continuous data streams, consumer-initiated interaction poses significant overhead in terms of number of request messages and in the delay until fresh data becomes available.
On the other hand, changing to a pure \emph{producer-initiated} interaction as favored by \ac{CEP} is problematic for many \ac{IoT} applications that build on the request-reply interactions. For example, applications like Amazon Alexa~\cite{Alexa2016} need personalized request-reply interaction.

In fact, \ac{ICN} ideally should offer efficient support for both interaction patterns as part of a unified communication model as illustrated in \Cref{fig:taxonomy}.
Initial work in the context of content routing has shown the potential of a unified communication model \cite{Carzaniga2011}.
However, performing CEP operations inside the network while efficiently realizing a unified communication model in \ac{ICN} is a challenge that we aim to address in this paper.
Thus, we present a novel \system architecture~\cite{INetCEPGithub2019} with the following contributions:
\begin{enumerate}
\item a unified communication layer to provide the functionality of \ac{CEP} at the network level,
\item a general \ac{CEP} query language that
specifies patterns for meaningful event detection over the \ac{ICN} substrate, in the form of \emph{query interests},
\item a query processing algorithm to resolve query interests, and
\item an open source implementation and evaluation of the proposed approach on a state-of-the-art \ac{ICN} architecture, \ac{NFN}, with two \ac{IoT} case studies.
\end{enumerate}

The paper is organized as follows. In Section \ref{sec:preliminaries}, we present preliminaries required to understand our approach. In Section \ref{sec:motivation}, we present two motivational \ac{IoT} use cases. In Section \ref{sec:probanddesign}, we describe the problem space and our system model. In Section \ref{sec:design}, we present the \system architecture, and in Section \ref{sec:eval}, we provide an evaluation. In Section \ref{sec:relatedwork}, we provide a comparison with related work. Finally, in Section \ref{sec:discussion}, we discuss possible extensions to our architecture before concluding in Section \ref{sec:conclusion}.

\section{Background}~\label{sec:preliminaries}

In this section, we briefly explain the building blocks of our work: \ac{CCN} that evolved into Named Data Networking (NDN) and Named Function Networking (NFN), and Complex Event Processing (CEP). \\

\textbf{Content-Centric Networking.}~\label{subsec:CCN} 
Jacobson et al.~\cite{Jacobson2009} proposed CCN\footnote{In the remainder of the paper, we will use the terms ICN and
CCN interchangeably.}, where communication is \emph{consumer-initiated}, consisting of two packets: \Interest  and \Data. 
A data object (payload of a \Data packet) satisfies an interest if the name in the \Interest packet is a prefix of the name in the \Data packet. 
Thus, when a packet arrives on a face\footnote{face stands for interface in \ac{CCN} terminology.} (identified by $face\_id$) of a CCN node, the longest prefix match is performed on the name and the data is returned based on a lookup. 

\textit{CCN data plane:} Each \ac{CCN} node maintains three major data structures: \ac{FIB}, \ac{CS} (also known as in-network cache), and \ac{PIT}. Once an \Interest arrives on a face, the node first checks its Content Store for a matching \Data packet by name. Upon a match, the \Data packet is sent via the same face it arrived from. Otherwise, the node continues its search in the \ac{PIT} that stores all the \Interest packets (along with its incoming and outgoing face) that are not satisfied. If an entry exists in the \ac{PIT}, the face is updated and the \Interest is discarded, because an \Interest packet has already been sent upstream. Otherwise, the node looks for a matching \ac{FIB} entry and forwards the \Interest to the potential source(s) of the data. 

\textit{NDN and NFN:} \ac{NDN}~\cite{Zhang2014a} emerged as a prominent architecture that builds on the principles of CCN's \emph{named data}. 
\ac{NFN} is another emerging architecture that focuses on addressing \emph{named functions} in addition to \emph{named data} by extending the principles of \ac{NDN}. 
NFN blends~\cite{Tschudin2014} data computations with network forwarding, by performing computational tasks across the CCN network. It represents \emph{named functions} on the data as Church's $\lambda-$calculus expressions that are the basis of functional programming. 
We aim to encapsulate CEP operators (cf. next section) as NFN named functions and hence resolve them in the network.
Yet, the proof-of-concept design of NFN focuses mainly on resolving functions on top of the CCN substrate. In contrast, we focus on continuous and discrete computations (push and pull), expressive representation of the computation tasks and their efficient distribution (cf. Section \ref{sec:design}). \\

\textbf{Complex Event Processing.}~\label{subsec:CEP}
CEP can process multiple online and unbounded data streams using compute units called \emph{operators} to deliver meaningful events to the consumers. The consumers specify interest in the form of a \emph{query} comprising of multiple operators.
 Some of the commonly used operators are defined in the following.
\begin{enumerate}
\item \texttt{\textbf{Filter ($\sigma$)}} checks a condition on the attribute of an event tuple and forwards the event if the condition is satisfied. 
\item \texttt{\textbf{Aggregate}} applies an aggregation function such as $max$, $min$, $count$, $sum$, $avg$, etc., on one or more event tuples. Hence, the data stream must be bounded to apply these operations. For this purpose, \emph{window} can be used.
\item \texttt{\textbf{Window}} limits the unbounded data stream to a window based on time or tuple size, such that operators like \texttt{Aggregate} can be applied on the selected set of tuples. 
\item \texttt{\textbf{Join ($\Join$)}} combines two data streams to one output stream based on a filter condition applied on a window of limited tuples.
\item \texttt{\textbf{Sequence ($\rightarrow$)}} detects causal or temporal relationship among two events applied on a window of selected event tuples from a data stream, e.g., if event $a$ caused event $b$ and event $a$ happened before event $b$.
\end{enumerate}
\begin{figure*}
	\centering
	\begin{subfigure}[t]{0.48\textwidth}
	\includegraphics[width=\linewidth]{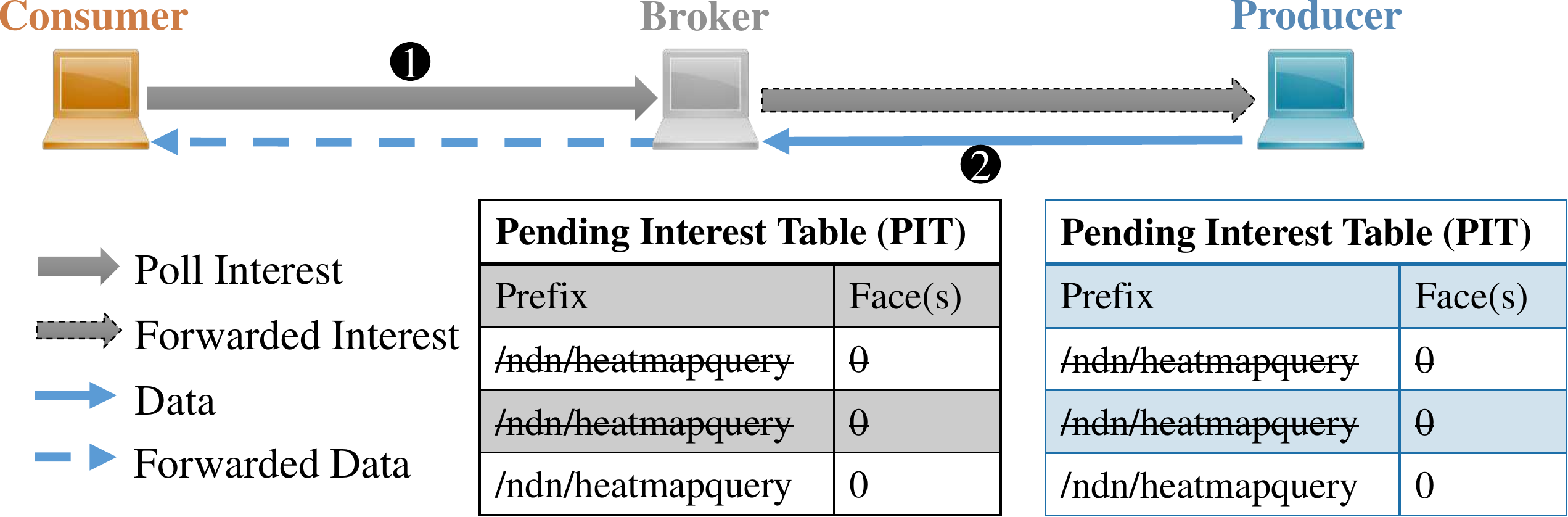}
	\caption{Limitation 1: Continuous polling leads to a lot of traffic and endless \ac{PIT} entries.}
	\label{fig:lim1}
	 \end{subfigure}%
	\hfill
    \begin{subfigure}[t]{0.48\textwidth}
	\includegraphics[width=\linewidth]{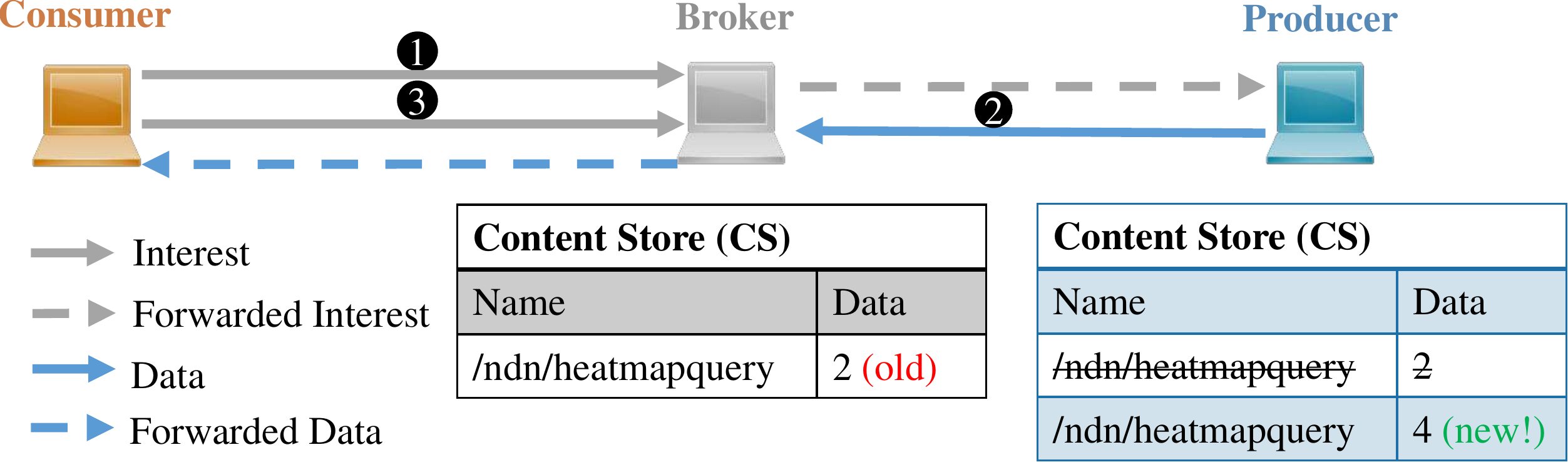}
	\caption{Limitation 2: Stale data due to an old \ac{CCN} \Interest.}
	\label{fig:lim2}
	\end{subfigure}
	\setlength{\abovecaptionskip}{5pt}
	\setlength{\belowcaptionskip}{-10pt}
	\caption{Limitations when standard consumer initiated communication is used in \ac{CCN} to support \ac{CEP}.}
	\label{fig:limconsumer}
\end{figure*}
These operators can be \emph{stateless} or \emph{stateful}. \texttt{{Filter}} and \texttt{{Aggregate}} are stateless operators, while the other operators are stateful and maintain the state of input tuples before emitting the complex event and therefore depend on multiple input tuples to be accumulated before actual emission.
\section{Motivating Use Cases} \label{sec:motivation}

\textbf{Use case I: Disaster Scenario.}
A natural disaster scenario is a prominent use case of \ac{ICN} architecture research.
It is drafted as one of the baseline scenarios in an active Internet Research Task Force (IRTF) draft of \ac{ICN} working groups~\cite{ICNIRTF2014}. 
A typical disaster management application is to generate a heat map showing live distributions of survivors in a disaster area~\cite{Koepp2014}. An important property of such an application is that the information must be updated continuously to delegate rescue workers to the hot spots and to monitor their operations, and hence is \emph{producer-initiated}. 
However, there are limitations of developing such an application using current \ac{ICN} architectures such as \ac{NDN} alone, due to the absence of support for \emph{producer-initiated} communication. \\

\textbf{Use case II: Internet of Things.}
We consider an \ac{IoT} application as the second use case for our approach since:
\begin{inparaenum}
\item it is one of the baseline scenarios for \ac{ICN} architectures~\cite{ICNIRTF2014} and 
\item \ac{IoT} traffic is among the most used type of traffic in the current Internet~\cite{IDC2017}. 
\end{inparaenum}
An intrinsic property of such applications is that \ac{IoT} devices produce continuous data streams, \eg sensor data that needs to be analysed, filtered, and derived to retrieve meaningful information for the end consumers, and hence it is oriented towards \emph{producer-initiated} interaction. 
One such application in the context of smart homes is short term load forecasting of energy consumed by smart plugs, which is useful, \eg for energy providers. The DEBS grand challenge 2014 focuses on this application. 

Although we target the above use cases in the following sections, our solution is not limited to these use cases, but the presented scenarios are representative to cover the design space of our solution. 

\section{Problem Space}\label{sec:probanddesign}
In this section, we first discuss the limitations of using straightforward solutions to motivate the need of our architecture (cf. \Cref{subsec:decisions}) and then we explain the system model of \system (cf. \Cref{sec:systemmodel}). 
\subsection{Design Challenges}\label{subsec:decisions}
We discuss the limitations of using straightforward solutions, \eg standard consumer-initiated, producer-initiated communication, or long lived \Interest packets for the purpose of supporting a wide variety of applications as also pointed here~\cite{Carzaniga2011}. Then, we study limitations of using \Interest packets to represent queries. Finally, we present limitations in performing operator graph processing at the consumer end.  

A straightforward solution to support \ac{CEP} is to use the standard consumer initiated communication of the \ac{CCN} architecture. We illustrate the problems using this naive solution in Figure~\ref{fig:limconsumer}. 
One way is that the consumers continuously issue \circled{1} a query at regular intervals, and \circled{2} the producer replies with the event of interest in a data (notification) packet. However, there are multiple problems with this solution, as indicated in~\Cref{fig:lim1}.

\begin{figure}[t]
	\centering
	\includegraphics[width=\linewidth]{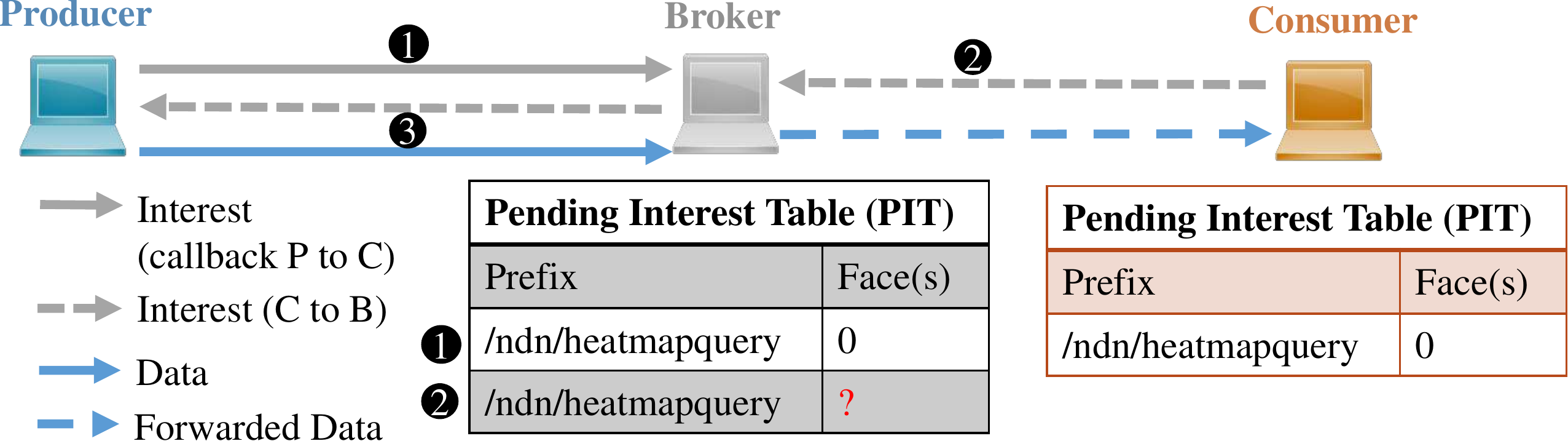}
	\caption{Limitation 3: Three way message and two kind of \Interest packets required.}
	\label{fig:lim3}
\end{figure}

\textbf{Limitation 1}: The continuous polling of a query by consumers generates a lot of overhead traffic and network state in the form of pending interests for only a few meaningful data packets. Each time a data packet is received, the pending interest is removed from \ac{PIT} since it is satisfied (represented as \st{strikethrough} in the figure). However, depending on the query interval a new entry is again created in the \ac{PIT} for each query. Also, the interval length of issuing query might determine the maximum latency at which the notification is delivered to the consumer, which might not be acceptable for latency sensitive applications, \eg autonomous cars. 

\textbf{Limitation 2} is to deal with the stale data in the cache or \ac{CS}, as represented in~\Cref{fig:lim2}. The consumers in a CEP application often need real-time updates on the latest data. For this reason, the query needs to be updated each time, otherwise it will retrieve the last cached \Data packet which is stale or obsolete in time. For instance, in ~\Cref{fig:lim2}, the broker still sends the old data to the consumer while the producer has generated a new data item for the query.
In addition, there should be a mechanism to expire the \Data packet at the right time, perhaps, immediately for the real-time updates. Solutions like appending sequence numbers (similar to TCP
) to the \Data packet can be applied. However, this will require additional synchronization mechanisms. 

Alternatively, another possibility is to support just producer-initiated transmission while using \ac{CCN} primitives (cf. \Cref{fig:lim3}). 
Although this is a viable option for some applications~\cite{Chen11}, it results in a three-way message exchange of what amounts to a one-way message. 
\circled{1} The producer sends an asynchronous \Interest packet that is not intended to fetch a \Data packet from the network but to announce the data \emph{name} and the callback from the consumer. 
\circled{2} The consumer then shows interest in the data \emph{name}, which is  
\circled{3} fulfilled by a \Data packet from the producer.

\begin{figure}[t]
	\centering
	\includegraphics[width=\linewidth]{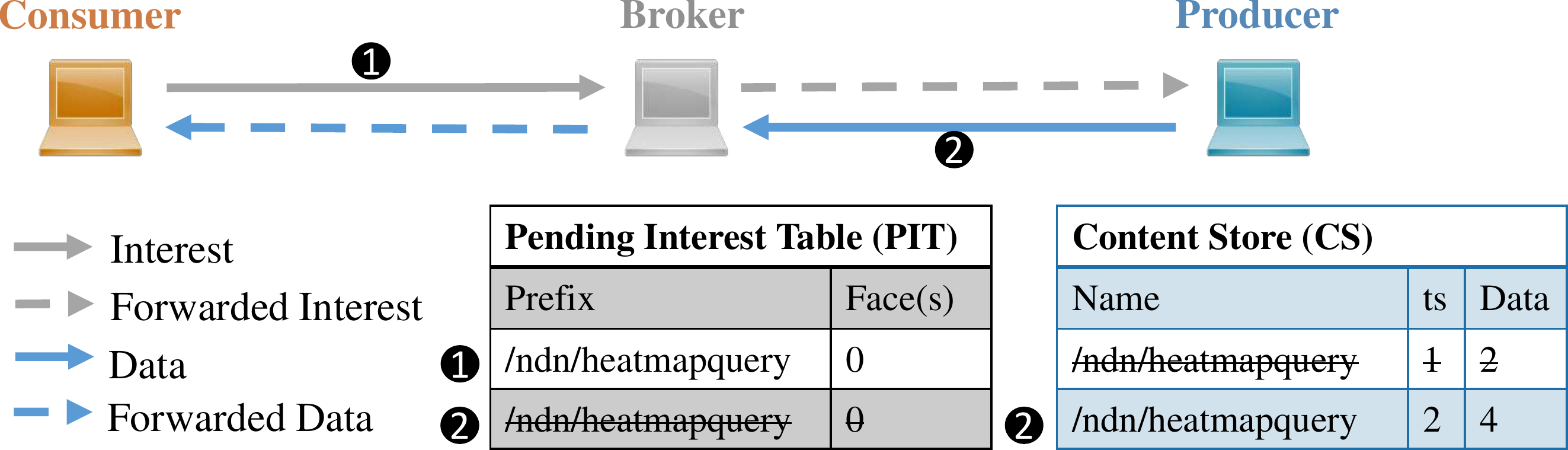}
	\setlength{\abovecaptionskip}{0pt}
	\caption{\system communication model supports pull-based communication and fetches latest \Data packet.}
	\label{fig:inetcep_pull}
\end{figure}

\textbf{Limitation 3}: Besides the overhead generated by a 3-way message, this design has other major issues. 
The interests leave an in-network state in the \ac{PIT} of \ac{CCN} nodes such that data can be fetched by the same path. \ac{CCN} typically performs a unicast of the \Interest packet, so that the \Data packet can follow the same path to the consumer. 
Such an application has to support two kinds of \Interest packets: a packet that is not supposed to fetch data, and a packet that is supposed to fetch data. 

\textbf{Limitation 4}:
Long-lived \Interest packets can be used in place of a \emph{query}, but this also has multiple side-effects. Similar to multiple interests, long-lived \Interest packets will also result in large in-network state (cf. \Cref{fig:lim1}). In addition, the long-lived \Interest packets will have to deal with stale data as explained earlier in Limitation 2 (cf. \Cref{fig:lim2}). 
To solve the aforementioned issues, we propose to have both \emph{consumer-initiated} and \emph{producer-initiated} interaction patterns coexisting under a unified \ac{CCN} communication layer. A \ac{CCN} architecture is unable to achieve this using existing packets and data structures as we saw above. Hence, we propose additional packets as a part of the communication model and handle them while processing CEP queries in the network 
as defined in Section \ref{subsec:unified}.


\textbf{Limitation 5:} CCN/NDN assumes a hierarchical naming scheme to address named data, \eg $/node/nodeA/temperature$, in order to fetch data objects \eg $35\degree C$, from the producers. A simple way to specify \ac{CEP} operations over data would be to represent this using the standard naming scheme, \eg a $min$ operator as \\
$/node/nodeA/min/temperature$. However, there are problems with this approach: 
\begin{inparaenum}
\item the \emph{name} cannot be used to correlate data from multiple producers, 
\item this would mean the processing is performed always at the consumer, which is inefficient and 
\item this is not extensible and not expressive, since adding more operators would mean appending them in the naming scheme, which reduces readability.
\end{inparaenum}

Hence, we need more than just \ac{CCN} \Interest packets that encapsulate name prefixes as stated above to represent CEP \emph{queries} over a \ac{CCN} network.
We propose  an \emph{expressive} query language that can correlate data from multiple producers and an efficient query parser to execute queries in the network (cf. Section \ref{subsec:language}).


\textbf{Limitation 6}: The \emph{query} specified by the consumers must be processed within the \ac{CCN} network. The \ac{CCN} resolution engine can resolve only \Interest packets to retrieve \Data packets based on the matching name prefix, but it cannot express \emph{query}.
A naive way to deal with this is to process the query at the consumer. However, this would overload the network with all the unnecessary data that could have been filtered on the way to the consumer and overload the consumer with all the processing. This  might result in a single node of failure, when the data becomes very \emph{big}. Thus, the processing needs to be performed in the \ac{CCN} network, \eg at the broker while being transmitted to the consumer. We provide this in two ways: (i) centralized query processing, where the entire query is processed at a single broker and (ii) distributed query processing, where the query operators are assigned to in-network nodes for processing (cf. Section \ref{subsec:ogprocessing}).

 \begin{figure}[t]
	\centering
	\includegraphics[width=\linewidth]{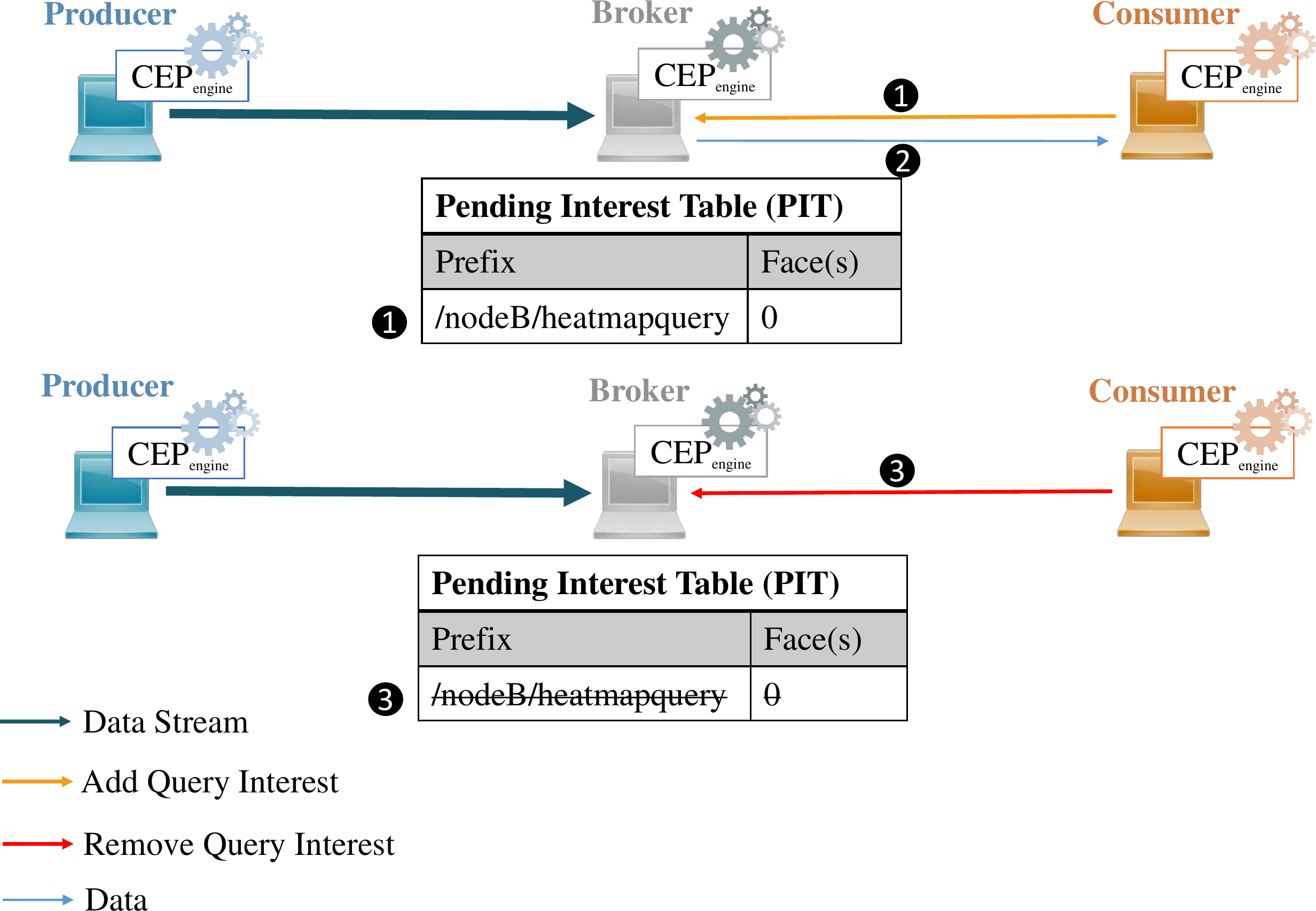}
	\setlength{\abovecaptionskip}{0pt}
	\caption{\system communication model supports push-based communication without creating endless \ac{PIT} entries.}
	\label{fig:inetcep_push}
\end{figure}

\subsection{\system System Model}\label{sec:systemmodel}
Every \ac{CCN} node can act either as a \emph{producer}, a \emph{consumer} or a \emph{broker}. Here, a broker is an in-network element, i.e., an \system aware \ac{CCN} router, while a producer or consumer is an end device, \eg a sensor or a mobile device. 
On the one hand, \circled{1} consumers can request a specific data item using an \Interest packet, where broker(s) forward(s) the request received by consumers to support \emph{anonymous request-reply communication}, as illustrated in \Cref{fig:inetcep_pull}. \circled{2} The producer replies with a data object contained in a \Data packet.
On the other hand, broker(s) process(es) the unbounded and ordered \emph{data streams} generated by producers to provide \emph{event-based communication}, which happens as illustrated in \Cref{fig:inetcep_push} and is explained below.

\begin{table*}
\footnotesize
\begin{tabular}{|p{2cm}|lp{3cm}|lp{7cm}|}
\hline
\multirow{2}{*}{\textbf{Characteristic}}            & \multicolumn{2}{c|}{\textbf{Old Architecture}}                                       & \multicolumn{2}{c|}{\textbf{Our Architecture}}                                                                     \\  
                                 & \ac{ICN} & Description                                     & \system           & Description                                                            \\ \hline
\multirow{5}{*}{\begin{minipage}{1.5cm}Packet Types (cf. \S\ref{subsubsec:handling})\end{minipage}}      & \Interest  & Consumer request                                & \Interest         & Consumer request                                                       \\  
                                 & \Data      & Producer reply                                  & \Data             & Producer reply                                                         \\  
                                 & -                         & -                                               & \DataStream       & Data stream of the form $<ts, a_1, a_2, .., a_m>$                          \\  
                                 & -                         & -                                               & \AddQInterest     & \ac{CEP} query subscription                           \\  
                                 & -                         & -                                               & \RmQInterest      & \ac{CEP} query unsubscription                         \\ \hline
\multirow{3}{*}{\begin{minipage}{2cm}Data Structures \\ (cf. \S\ref{subsubsec:components})\end{minipage}} & \ac{PIT} & Stores pending interests of consumer            & \ac{PIT}        & Stores pending interests and query interests                           \\  
                                 & \ac{CS}  & Stores data packets                             & \ac{CS}         & Stores dat
                                 a packets and buffers the data stream for stateful operators \\  
                                 & \ac{FIB} & Stores forwarding information towards producers & \ac{FIB}        & Stores forwarding information towards producers and consumers interested in CEP query                      \\ \hline
{\begin{minipage}{2cm}Data Processing  \\(cf. \S\ref{subsec:language}, \S\ref{subsec:ogprocessing})  \end{minipage}}                & -                         & -                                               & \ac{CEP} engine & Parse, process and derive complex events                               \\ \hline
\end{tabular}
\setlength{\abovecaptionskip}{0pt}
\setlength{\belowcaptionskip}{-15pt}
\caption{Description of differences in traditional \ac{ICN} vs \system architecture ("-" means no support).}
\label{tab:differences}
\end{table*}

A producer multicasts the data stream (\DataStream packet) towards the broker network, which disseminates the stream all over the network (push). The \DataStream packet is forwarded to further brokers in the network if there are consumers downstream for \emph{query interest} ($qi$). An efficient event dissemination can be achieved by using routing algorithms, \eg defined in this work~\cite{Baldoni2007}, by looking at the similarity score of the $qi$.
\circled{1} A consumer issues a query by sending an \AddQInterest packet comprising $qi$ (top \Cref{fig:inetcep_push}). Each $qi$ encapsulates a \ac{CEP} query $q$ that is processed by interconnected brokers in $B$ forming a \emph{broker network}. \circled{2} The $qi$ is stored in the \ac{PIT} of the receiving broker until a \circled{3} \RmQInterest packet is received that triggers the removal of $qi$ from the \ac{PIT} (bottom \Cref{fig:inetcep_push}). 
Unlike a conventional \ac{CEP} system, the event-based communication in the \system happens in the underlay CCN network.

The query $q$ induces a directed acyclic operator graph $G$
, where a vertex is an operator $\omega \in \Omega$ and an edge represents the data flow of the data stream $D$. Each operator $\omega$ dictates a processing logic $f_\omega$. We explain the constituents: the communication model, the query model, and the operator graph model below.
 \newline

\textbf{Communication Model.}\label{subsec:comm_model}
We provide five types of packets to support both kinds of interaction patterns. The 
\Interest (\emph{request}) packet is equivalent to CCN's Interest packet that is used by the consumer to specify interest in any named data or named function.
The \Data (\emph{reply}) packet is a \ac{CCN} data packet that satisfies an Interest. It also encapsulates the \emph{complex event} ($ce$) as described later.
The \DataStream packet represents a data stream of the form $<ts, a_1, \ldots, a_m>$. 
Here, $ts$ is the time at which a tuple is generated and $a_i$ are the attributes of the tuple.
The \AddQInterest packet represents the event of interest in the form of a \ac{CEP} query $q$.
The \RmQInterest packet represents the \ac{CEP} query that must to be removed for the respective consumer (so that it no longer receives complex events). 
The \ac{CCN} forwarding engine (data plane) is enhanced to handle these packets, as below. 
\newline

\textbf{Query Model.} \label{subsec:query_model}
The \system query language is based on two main design goals: 
    it should deal with both pull (data from relations) and push (time series data streams) kind of traffic, and
    support standard \ac{CEP} operators~
  (as identified in Section \ref{subsec:CEP}) over the \ac{CCN} data plane.

Thus, a query ($q$) must be able to 
   capture time series data streams as well as relations of the form $<ts, a_1, .., a_m>$ and 
   define an operator $\omega$ with processing logic $f_\omega$ in a way that it is extensible. 
\newline

\textbf{Operator Graph Model.} \label{subsec:op_model}
The operator graph $G$ is a directed acyclic graph of \emph{plan nodes}. The vertex of the graph is a plan node that encapsulates a single operator ($\omega$), while the links between plan nodes represents the data flow from the bottom of the graph to the top. The operator graph can be processed centrally or collaboratively in a distributed manner by mapping it to the underlay \ac{CCN} network. 
In distributed \ac{CEP}, typically, an operator placement mechanism defines a mapping of an operator graph $G$ onto a set of brokers, to collaboratively process the query. The placement needs to be coordinated with the forwarding decisions for efficient processing over the \ac{CCN} data plane.

\section{\system architecture} \label{sec:design}
We identify the following three broad requirements for the \system architecture from our discussion in the previous section. 
\\
\textbf{R1} A unified communication layer supporting both \emph{producer} and \emph{consumer} initiated communication (cf. Section \ref{subsec:unified}).  \\
\textbf{R2} An expressive \ac{CEP} query language for specifying the event of interest (cf. Section \ref{subsec:language}). \\
\textbf{R3} Resolution of \ac{CEP} queries by efficient and scalable in-network query processing (cf. Section \ref{subsec:ogprocessing}). 

We address each of these requirements below.

\subsection{Unified Communication Layer} \label{subsec:unified}

In this section, we explain the extension of the \ac{CCN} data plane to enable \ac{CEP}.
In our approach, each \ac{CCN} node $n \in N$ maintains a Content Store or cache (\ac{CS}), a Pending Interest Table (\ac{PIT}), a Forwarding Information Base (\ac{FIB})
 and a \ac{CEP} engine. In the following, we explain the function of these main building blocks and the data plane handling. Subsequently, we detail on the newly introduced packets \AddQInterest, \RmQInterest and \DataStream and simultaneously the solution to the limitations identified in \Cref{subsec:decisions}.

\begin{figure}
    \centering
    \includegraphics[width=\linewidth]{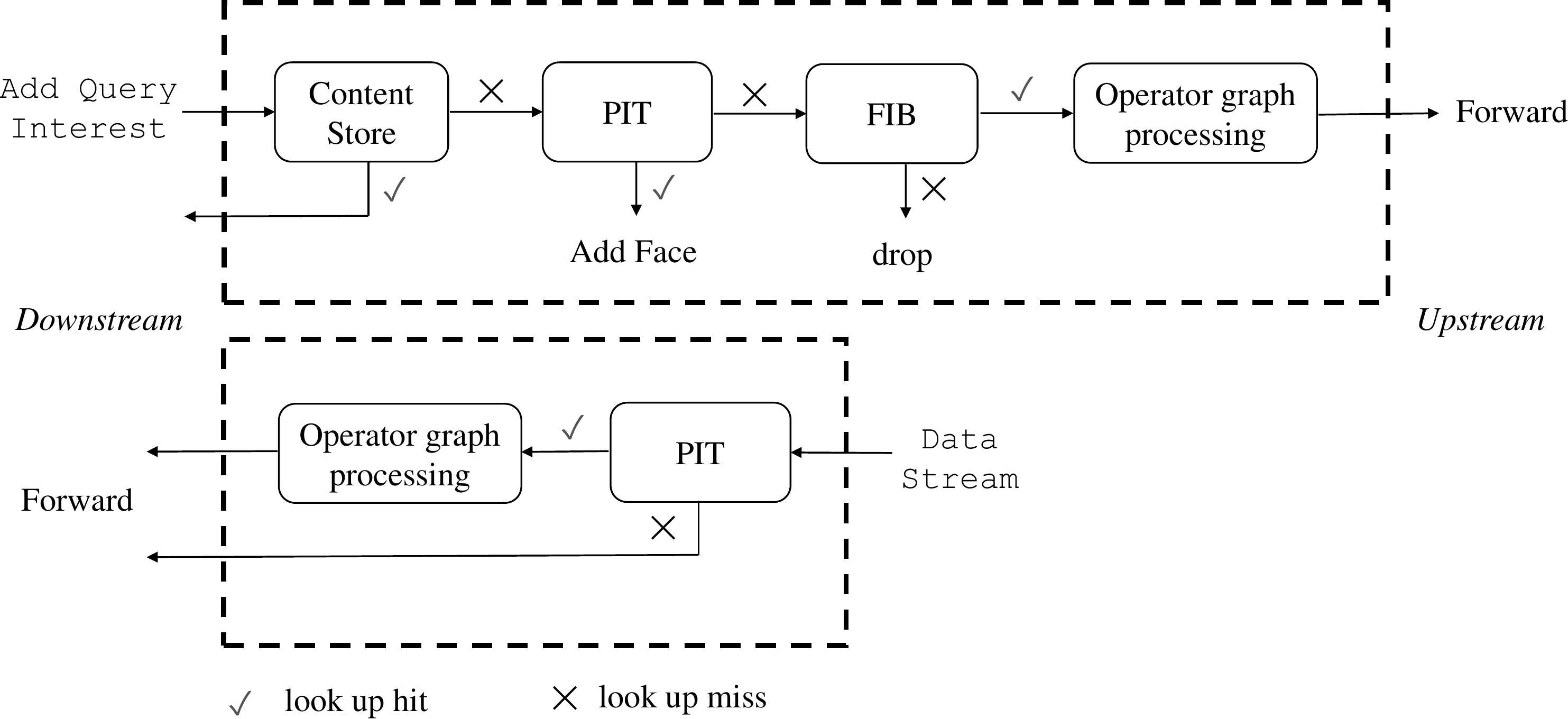}
    \setlength{\abovecaptionskip}{-5pt}
    \setlength{\belowcaptionskip}{0pt}
    \caption{High level view of packet handling in \system architecture.}
    \label{fig:overall}
\end{figure}
\subsubsection{Node Components}~\label{subsubsec:components}
The \ac{CS} stores all the data objects associated with the \ac{CCN} \Interest, additionally, the time-stamped data objects associated with the \emph{query interests ($qi$)}. 
The data object returns a value associated with the $qi$, \eg if the $qi$ is the sum of victims in a disaster location, then the data object contains value 20. Hence, if multiple consumers are interested in the same $qi$, the query is not reprocessed but the data is fetched directly from the \ac{CS}.
To deal with \textbf{Limitation 2} of stale cache entries for $qi$ (cf. Section \ref{subsec:decisions}), we store logical timestamps along with cache entries while storing $qi$ in the \ac{CS}. Hence, always up to-date data objects are stored, while old entries are discarded (cf. \Cref{fig:inetcep_pull}).

The \ac{PIT} stores the pending $qi$ so that the \emph{complex event} ($ce$) could follow the path, i.e., the in-network state created in \ac{PIT} to the consumer. 
The $\mi{face}$ information is also stored in the \ac{PIT} entry to keep track of $qi$'s consumers. In contrast to consumer-initiated interaction, the $ce$ must be notified to the interested consumers as and when detected in real-time. 
Thus, as new data is received, the \emph{query interests} in the \ac{PIT} are re-evaluated as explained later in Algorithm~\ref{algo:addqianddatastream}. 
In the former, we are referring to only continuous \DataStream packet since we also support fetching data from relations (this is handled conventionally using a \ac{CCN} \Interest packet as explained in Section \ref{sec:preliminaries}).

The reasons why we distinguish between \AddQInterest ($qi$) and CCN \Interest packets are: (i) $qi$ is invoked on receipt of \AddQInterest as well as the \DataStream packet, (ii) removal of the \ac{PIT} entry is not based on a \Data packet retrieval but on the reception of the \RmQInterest packet and (iii) $qi$ retrieves \Data packets asynchronously.  
In summary, we deal with \textbf{Limitation 3} and \textbf{Limitation 4} by asynchronously handling the $qi$ instead by 3-way message exchange and efficiently managing \ac{PIT} entries, respectively. 
We deal with \textbf{Limitation 1} by storing $qi$ in \ac{PIT} and asynchronously delivering event notifications to the consumers.

The \ac{FIB} table gets populated as the producer multicasts to the broker network leaving a trail to the data source. In this way the data processing is performed efficiently along the path from producer and consumer.
Finally, the \ac{CEP} engine holds the processing logic $f_\omega$ for each operator $\omega$ and is responsible for parsing, processing, and returning the result to the next node towards consumer (cf. Sections \ref{subsec:language} and \ref{subsec:ogprocessing}).

\IncMargin{1em}
\begin{algorithm}
\footnotesize
\KwVar{$\mi{CS} \leftarrow \text{content store of current node}$ \\
$\mi{PIT} \leftarrow \text{pending interest table of current node}$ \\
$\mi{FIB} \leftarrow \text{forwarding information base}$\\
$\mi{qi} \leftarrow \text{requested query interest}$ \\
$\mi{result} \leftarrow \text{query result}$ \\
$\mi{facelist} \leftarrow \text{list of all faces in PIT}$\\
$\texttt{DataStream} \leftarrow \text{\DataStream packet}$\\
$\mi{data} \leftarrow \text{data that resolves the qi}$
}
\BlankLine
\Function{$\textsc{AddQueryInterest}(\mi{qi})$}
{ \label{algline:addqistart}
	
	\If{ $qi$ is found in  $\mi{CS.\textsc{lookup}(qi)}$} { \label{algline:csstart}
		$ \mi{data} \leftarrow \mi{CS}.\textsc{fetchContent}(\mi{qi}) $\;
		$ \text{return } \mi{data}$\; (Discard \texttt{AddQueryInterest}) \label{algline:csstop}
	}
	\ElseIf { $qi \text{ found in } \mi{PIT.\textsc{lookup}(qi)} $} { \label{algline:pitstart}
		$\textsc{ProcessQiInPIT}(\mi{qi}, \texttt{AddQueryInterest})$		\label{algline:pitstop}
	}
	\ElseIf { $qi \text{ found in } \mi{FIB.\textsc{lookup}(qi)} $} { \label{algline:createpitentry}
	 	$\mi{\textsc{CreateOperatorGraph}(\mi{qi})}$ (Refer Algorithm \ref{algo:parser})\;
   		Forward \texttt{AddQueryInterest}\; \label{algline:addqistop}
	}
	\Else {
	    (Discard \texttt{AddQueryInterest}) \label{algline:addqistop}}
}
\Function{$\textsc{ProcessDataStream}(\texttt{DataStream})$} 
{ \label{algline:dsstart}	
	\ForEach{ $\mi{qi} \in PIT$ } 
	{
		\If { \texttt{DataStream} \text{ satisfies } $qi$ } {
			$\textsc{ProcessQiInPIT}(\mi{qi}, \texttt{DataStream})$	\; \label{algline:processPITDS}
		}
		\Else {
			Forward \texttt{DataStream}\;  \label{algline:dsstop}
		}
	}	
}	
\Function{$\textsc{ProcessQiInPIT}(\mi{qi}, \mi{packet})$} 
{	\label{algline:qiinpitstart}
	\If{$\mi{packet} == \texttt{DataStream} \text{ and } \mi{packet}.\mi{ts} > \mi{qi.ts}$} { \label{algline:checkifdsstart}  
		$\mi{\textsc{CreateOperatorGraph}(\mi{qi})}$ (Refer Algorithm \ref{algo:parser}) \;
		Forward $\mi{packet}$\; \label{algline:checkifdsend}
	}
	\Else {	
		$\mi{facelist} \leftarrow \mi{PIT}.\textsc{getFaces}(\mi{qi}) $\; \label{algline:facelist}
		\If { $qi.\mi{face}$ is not found in $\mi{facelist}$} {  \label{algline:notfoundinpit}
			$ \mi{PIT}.\textsc{addFace}(\mi{qi}) $\; \label{algline:discardinpit} 
			}
		Discard $\mi{packet}$\;	 \label{algline:qiinputstop}				
	}
}

\setlength{\belowcaptionskip}{-15pt}
\caption{\AddQInterest and \DataStream packet handling.}
\label{algo:addqianddatastream}
\end{algorithm}
\subsubsection{Data Plane Handling}\label{subsubsec:handling}

In Algorithm~\ref{algo:addqianddatastream} (lines~\ref{algline:addqistart}-\ref{algline:addqistop}) and \Cref{fig:overall}, we define the handling of \AddQInterest and \DataStream packets at the broker end in a \ac{CCN} network.
The processing of $qi$ stored in \ac{PIT} is triggered based on the receipt of these two packets as follows: 
\begin{inparaenum}
\item when an \AddQInterest packet is received at a broker (line~\ref{algline:addqistart}) and
\item due to the continuous arrival of new \DataStream packets (line~\ref{algline:dsstart}).
\end{inparaenum}
This is in contrast to the \ac{PIT} entry of CCN \Interest, which is checked only on the receipt of a new \Interest packet.

When an \AddQInterest arrives, the broker checks if the (up to-date) data object corresponding to the $qi$ exists in the \ac{CS}. If this is true, the broker forwards the data object to the consumer and discards the $qi$ (lines~\ref{algline:csstart}-\ref{algline:csstop}). This is because the $qi$ is already processed at one or more brokers and a matching \Data packet (with latest timestamp) is found in the CS or cache. The resolution to the \Data packet is explained later in Section \ref{subsec:language}.  If the cache entry is not found, the broker continues its search in the \ac{PIT} table (lines~\ref{algline:pitstart}-\ref{algline:pitstop}). 

If $qi$ is found in \ac{PIT} 
and the face corresponding to the query interest does not exist (lines~\ref{algline:facelist}-\ref{algline:discardinpit}), a new ${face\_id}$ (from which the interest is received) is added. Conversely, if the face entry is found in \ac{PIT}, this means the $qi$ is being processed and hence the packet is discarded (lines~\ref{algline:notfoundinpit}-\ref{algline:qiinputstop}). 
However, if no entry in \ac{PIT} exists, this means that the consumer's interest reaches first time at the broker
network. Therefore, a new entry for $qi$ is created and the $qi$ is processed by first generating an operator graph (cf. Section \ref{subsec:language}) and then processing it (cf. Section \ref{subsec:ogprocessing}) (lines~\ref{algline:createpitentry}-\ref{algline:addqistop}). 
 
A \DataStream packet is also handled similarly to the \AddQInterest packet (lines~\ref{algline:dsstart}-\ref{algline:dsstop}), except for the fact that the query processing is triggered if \DataStream satisfies a query interest in \ac{PIT} and it is a new packet  (lines~\ref{algline:checkifdsstart}-\ref{algline:checkifdsend}). This means that the $qi$ performs an operation on the data object contained in the \DataStream packet. In this case, query processing is triggered because it may contribute to the generation of a new $ce$. 
In addition, if the broker does not have a matching $qi$ in \ac{PIT} entry, this means it is not allocated to operator graph processing and hence it is forwarded to the next broker (line~\ref{algline:dsstop}). The \DataStream is forwarded if there are consumers downstream by looking at the \ac{FIB} entry. 

When a \RmQInterest packet is received at a broker, the node looks up its \ac{PIT} table for an entry of the $qi$. If found, it removes the \ac{PIT} entry for $qi$ and the \RmQInterest packet is forwarded to the next node. 
It is done in a similar way as the \ac{PIT} entry corresponding to a \ac{CCN} \Interest packet is removed when a matching \Data packet is found.

To summarize, in~\Cref{tab:differences}, we show the differences of the \system architecture in comparison to the standard \ac{CCN} architecture in terms of the packet types, the data plane, and the processing engine. 
We show that with minimum changes in the data plane, we support both \emph{consumer-} and \emph{producer-}initiated traffic. 

\subsection{A General CEP Query Language} \label{subsec:language}
In this section, we present a general \ac{CEP} language to resolve \textbf{Limitation 5}. 
By doing this, we provide a means to resolve \ac{CEP} queries expressed as $qi$ (query interests) on the data plane of \ac{CCN}. The grammar definition of the query language can be found in \Cref{apndix:grammar}. 
We aim for three main design goals for the query language and the parser: 
\begin{inparaenum}
\item distinguishing between pull and push based traffic,
\item translating a query to an equivalent name prefix of the \ac{CCN} architecture, and 
\item supporting conventional relational algebraic operators and being extensible such that additional operators can be integrated with minimum changes. This is to ensure easy integration of existing and new \ac{IoT} applications.
\end{inparaenum}
We provide the definition of \system query language in Section \ref{subsubsec:definition} and the parser in Section \ref{subsubsec:parser}. 

\subsubsection{Query Language} \label{subsubsec:definition}
Each operator in a query behaves differently based on the input source type, i.e., consumer- and producer-initiated interaction, which is done based on the reception of a \Data packet or a \DataStream packet, respectively.
The \Data packet is processed and returned as a data object, as conventionally done in the \ac{CCN} architecture. For instance, a \texttt{Join} operator placed on broker $C$ can join two data objects, $<lat1, long1>$ with name prefix $/node/A$ and $<lat2, long2>$ with $/node/B$ to produce $<lat1, long1, lat2, long2>$ with $/node/C$.

 \begin{figure}
 	\centering
	\includegraphics[height=3cm]{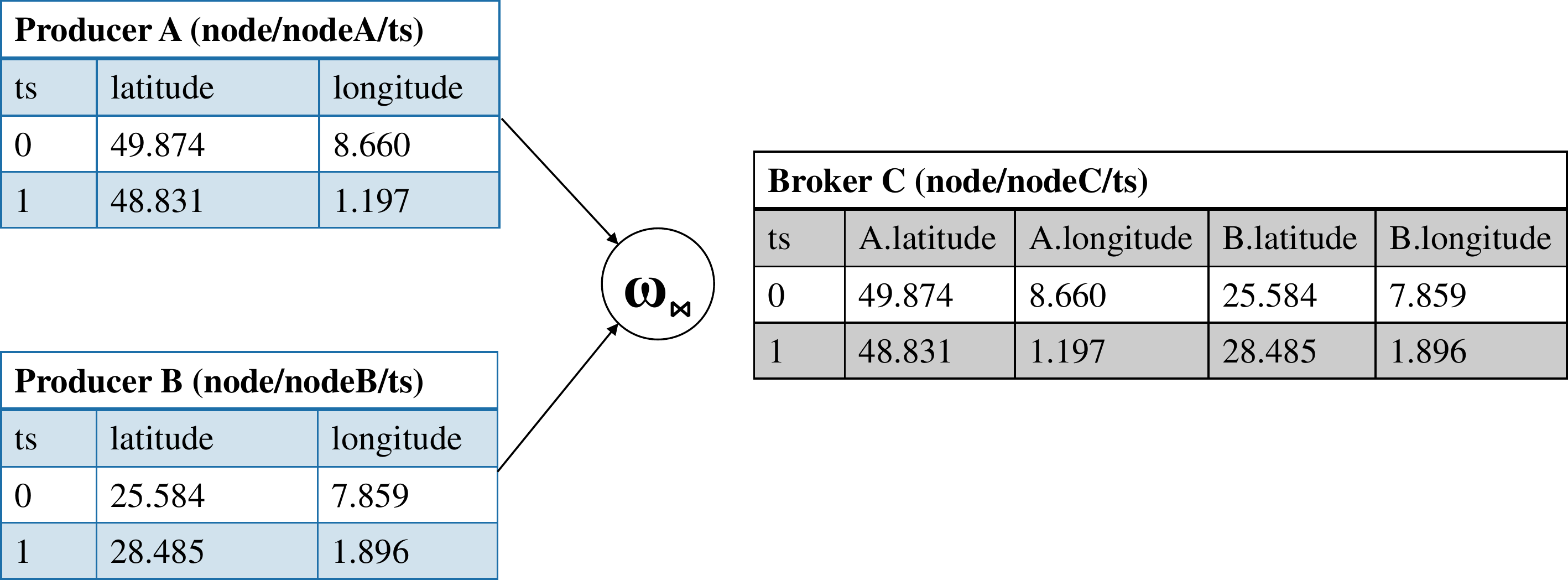}
	\caption{Join of two data stream packets with window size of 2s in a CCN network.}
	\label{fig:icn_join} 
 \end{figure}

In contrast, a \DataStream is processed and transformed either into an output stream (another \DataStream packet), or can be transformed to derive a \Data packet, containing, \eg a \texttt{boolean} variable depending on the \ac{CEP} query. 
For instance, a join of two continuous data streams expressing location attributes of producer A with location attributes of producer B leads to the generation of a new data stream, as illustrated by broker C in  \Cref{fig:icn_join}.

We express the standard CEP operators as explained in \Cref{sec:preliminaries} using the \system language below\footnote{We have implemented all standard CEP operators defined in Section \ref{sec:preliminaries} including \texttt{SEQUENCE} operator but only present the representative operators relevant for our use cases.}.

\begin{lstlisting} [caption={Selects a sliding window of tuples for 4s from gps source 1.}, label={query:window}, captionpos=b, abovecaptionskip=0pt, belowcaptionskip=-5pt]
WINDOW(GPS_S1, 4s)
\end{lstlisting}

\begin{lstlisting} [caption={Selects tuples with latitude value less than 50 pts from window Query~\ref{query:window}.}, label={query:filter}, captionpos=b, belowcaptionskip=-5pt]
FILTER(WINDOW(GPS_S1, 4s),'latitude'<50)
\end{lstlisting}

\begin{lstlisting} [caption={Performs join of the two resulting tuples from Query~\ref{query:filter} with gps source 1 and 2.}, label=query:join,captionpos=b, belowcaptionskip=-5pt]
JOIN(
  FILTER(WINDOW(GPS_S1, 4s), 'latitude'<50), 
  FILTER(WINDOW(GPS_S2, 4s), 'latitude'<50), 
  GPS_S1.'ts' = GPS_S2.'ts'
) 
\end{lstlisting}

The stateful operators \eg \texttt{Window} and \texttt{JOIN} must store the accumulated tuples in some form of a  readily available storage. For this, we make use of in-network cache, the \ac{CS}, that readily provides data for the window operator. This can be highly beneficial, \eg in a dynamic environment where state migration is necessary.

The \system query language provides an \emph{abstract}, \emph{simple}, and \emph{expressive} external DSL that translates the \ac{CEP} query to equivalent \ac{NFN} lambda ($\lambda$) expressions, as explained later in \Cref{subsubsec:parser}. The \system language abstracts over the complexity of lambda expressions (as seen in equation below), so that CEP developers can easily perform data plane query processing on an \ac{ICN} network. For instance, a general $\lambda$ expression of a \texttt{JOIN} query defined in Query~\ref{query:join} is given as follows. 

\begin{lstlisting} 
(call <no_of_params> /node/nodeQuery/nfn_service_Join 
(call <no_of_params> /node/nodeQuery/nfn_service_Filter 
(call <no_of_params> /node/nodeQuery/nfn_service_Window
 4s), 'latitude' < 50) (call <no_of_params> 
/node/nodeQuery/nfn_service_Filter 
(call <no_of_params> /node/nodeQuery/nfn_service_Window
 4s), 'latitude' < 50) GPS_S1.'ts' = GPS_S2.'ts')
\end{lstlisting}

\IncMargin{1em}
\begin{algorithm}[t]
\scriptsize
\KwVar{$query \leftarrow$ the input CEP query \\
		$\tau\mi{curList} \leftarrow$ top down list of 3 $\omega$ of tuple $\tau$ \\
		$\omega_{cur} \leftarrow$ current operator }
\BlankLine

\Function{$\textsc{createOperatorGraph} (query)$}
{ \label{algoline:createG}
	$\tau\mi{curList} \leftarrow$  \textsc{getCurList}($query$)\;
	\textsc{parseQuery($\tau\mi{curList})$}\; \label{algoline:stopG}
}

\Function{$\textsc{parseQuery}(\tau\mi{curList})$}
{ \label{algoline:startparsing}
	$\omega_{cur} \leftarrow $\textsc{getOperator}$(\tau\mi{curList})$\; 
	$\mi{nfnExp}  \leftarrow \textsc{constructNFNQuery}(\omega_{cur})$\;
	$node \leftarrow \text{new } \textsc{Node}(\mi{nfnExp})$\;

	\If{$size(\tau\mi{curList}) == 1$}{
		return $node$\;
	}
	\ElseIf{$size(\tau\mi{curList}) > 1$}{
	
		\textsc{parseQuery}($\tau\mi{curList}.left)$\;
  		\textsc{parseQuery}($\tau\mi{curList}.right)$\;
		return $node$\; 	\label{algoline:stopparsing}    	
    }
    
}

\caption{Recursively generating the operator graph}
\label{algo:parser}
\end{algorithm}
\DecMargin{1em} 

Here, \texttt{<no\_of\_params>} is the number of parameters in the $\lambda$ expression, \texttt{nfn\_service\_Join} is the name of the operator (join operator) in the query, $4s$ is the window size and the remaining are the filter and join conditions, respectively. Each operator is preceded by \texttt{/node/nodeQuery/..} which represents the \textit{name} of the node that is used to place the operator \eg nodeA. This is done at runtime by placing \texttt{nodeQuery} on the node name selected by the operator placement algorithm (cf. Section \ref{subsec:ogprocessing}) to process the operator in a centralized or a distributed manner. The translation of a \ac{CEP} query to the above $\lambda$ expression is discussed in the next section.

\subsubsection{Query Parser}\label{subsubsec:parser}
In Algorithm~\ref{algo:parser}, we express the \system query parser as a recursive algorithm to map the query (\eg Query~\ref{query:join}) to generate an equivalent \ac{NFN}'s \emph{ $\lambda$ expression} (\eg given above). 
A CEP query is transformed into an operator graph $G$ (lines~\ref{algoline:createG}-\ref{algoline:stopG}), which is a binary graph tree defined as a tuple $\tau = (L, S, R)$. Here, $L$ and $R$ are binary trees or an empty set and $S$ is a singleton set, \eg a single operator ($\omega$). The query parser starts parsing the query in a specific order, i.e., in a top-down fashion that marks the dependency of operators as well. This implies each leaf operator is dependent on its parent. Thus, the parser starts by iterating top down the binary tree starting from the root operator $\omega_{cur}$ (line~\ref{algoline:stopG}), where $cur = root$ in the first step. The traversal is performed in a depth-first pre-order manner (visit parent first, then left (L) and then right (R) children) (lines~\ref{algoline:startparsing}-\ref{algoline:stopparsing}). 

\begin{figure}[t]
\includegraphics[width=\linewidth]{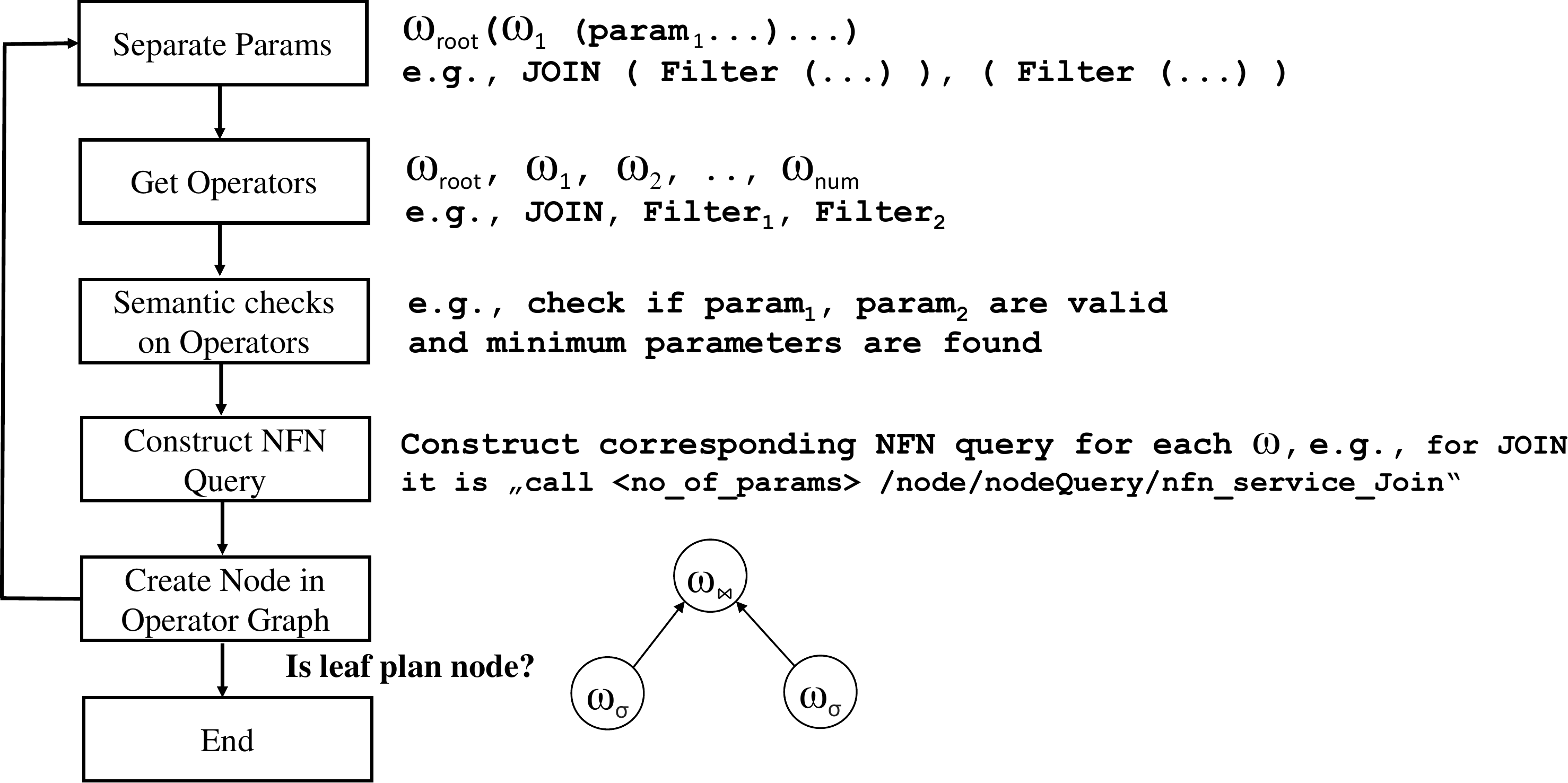}
\setlength{\abovecaptionskip}{-5pt}
\caption{Query parser workflow based on Algorithm~\ref{algo:parser}.}
\label{fig:parserworkflow}
\end{figure}

The workflow of the query parser algorithm for an example query of the form of Query~\ref{query:join}, is illustrated in Fig. \ref{fig:parserworkflow} and explained in the following.
We start by extracting the operator name ($\omega$) by separating the parameters into a list. 
 We create a logical operator graph for each query by instantiating the operators and their data flow.
 An operator is created only after the semantic checks on the operator are verified, e.g., if the $\omega$ is valid, and/or it has valid parameters. 
 Once all the semantic checks are verified, we continue processing the query recursively as in Algorithm~\ref{algo:parser}, by generating the corresponding \ac{NFN} query and
 creating logical \emph{plan nodes} for operators that are assigned to the broker network for processing the operator graph. 

\subsection{Operator Graph Processing} \label{subsec:ogprocessing}
This module processes the query either centrally or in a distributed way by mapping the operator graph plan nodes to the brokers ($B$) or \system aware \ac{CCN} routers after operator graph construction (cf. \Cref{subsec:language}). 
This is because central processing might not be sufficient for all the use cases, \eg when the amount of resources required to process the queries increases with the number of operators and/or queries. 
It is, therefore, necessary to distribute operators on multiple brokers. 
In this way, the network is also not unreasonably loaded by the queries, while the network forwarding is not disturbed.

In~\Cref{fig:ceptimeline}, we show the timeline of distributed query processing. 
In case of central processing, only parsing and deployment is required. 
\circled{1}  The broker that first receives the $qi$ from the consumer parses the \ac{CEP} query and forms an operator graph (as described in \Cref{subsubsec:parser}).
This broker becomes the placement coordinator and coordinates the further actions taken for operator graph processing. 
\circled{2} The coordinator builds the path where the operator graph is processed based on a criteria, \eg minimum latency and selects other broker nodes for operator placement. 
This is along the path from producer towards the consumer. 
It is important because the \Data packets are forwarded as well as processed along this path (in-network processing). 
\circled{3} The coordinator recursively traverses the operator graph, while assigning the \ac{CEP} operators or translated \emph{named functions} (cf.~\Cref{subsec:language}) to the \ac{CCN} routers.
The resulting $ce$ is encapsulated in a \DataStream or a \Data packet, which is received at the root node of the operator graph and forwarded to the consumer. 

In the following, we describe the collection of the monitoring information related to the \ac{CCN} nodes by the placement coordinator (cf.~\Cref{subsubsec:nwdiscovery}) and the assignment of operators based on this information  by the placement module (cf.~\Cref{subsubsec:placement}). 
In principle, the role of the coordinator is decided by the concrete operator placement algorithm. 
\system supports different means of coordination and hence operator placement algorithms. 
Yet, for the algorithm defined below, the node where the first (or root) operator is deployed is the coordinator. 

\subsubsection{Network Discovery Service} \label{subsubsec:nwdiscovery}
\begin{figure}[t]
\includegraphics[width=\linewidth]{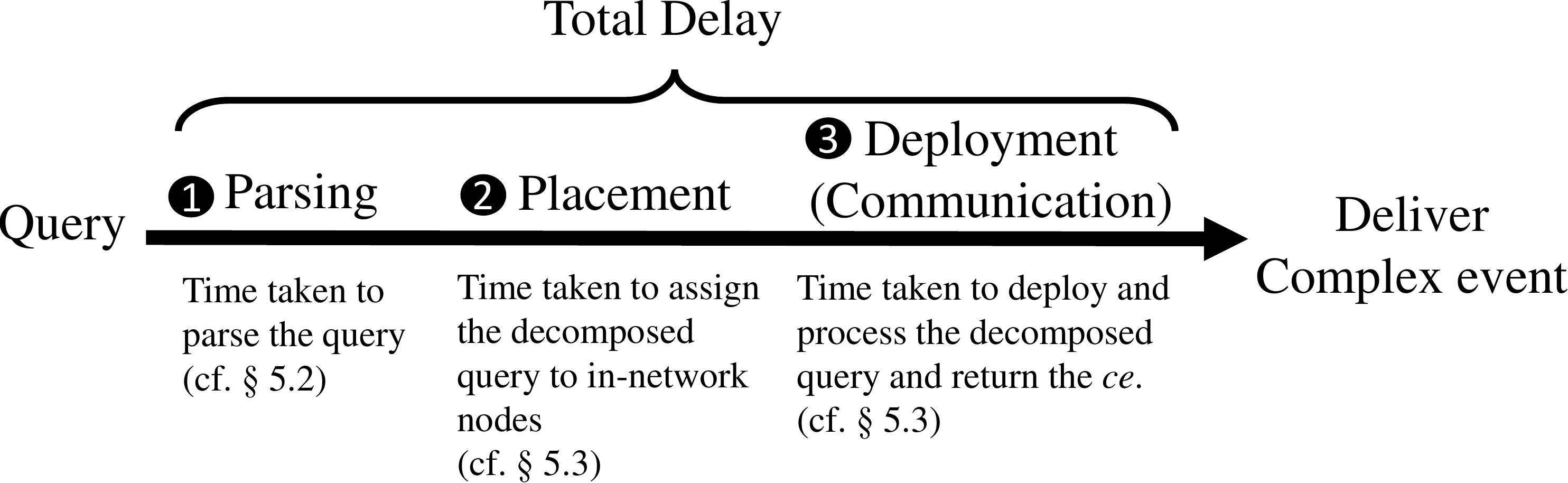}
\setlength{\abovecaptionskip}{-5pt}
\caption{Timeline of distributed query processing.}
\label{fig:ceptimeline}
\end{figure}

The placement coordinator fetches and maintains the monitoring information related to the node or network to place the operators on the right set of brokers or a single broker. 
Since different \ac{CEP} applications might be interested in optimizing distinct \ac{QoS} metrics, the network discovery service can be updated accordingly to monitor the respective metric(s). 
At the moment, we provide monitoring for the end-to-end delay which is important for our representative use cases. The end-to-end delay is defined as the complete timeline as illustrated in \Cref{fig:ceptimeline} from query parsing to the delivery of the complex event.

The node and network information is retrieved as a \texttt{Data} packet with name prefix, \eg\emph{/node/node\_id/delay} only on fetch basis (whenever required). 
The placement coordinator subscribes for this information and hence maintains the global (centralized) or local (decentralized) knowledge on the network. 
The cluster coordinators can be elected for decentralized placement as dictated in the placement literature~\cite{Lakshmanan2008}. 
By looking at the node and network characteristics, \eg average delay, the placement coordinator selects one or more nodes for operator placement (defined next).

\subsubsection{Operator Placement Module} \label{subsubsec:placement}
The operator placement module handles distributed query processing in case the processing requests, \eg in terms of query interests, exceed the network or node capacity. 
It works in conjunction with the placement coordinator, which is a primary component to provide operator placement decisions. 
This module is responsible for
\begin{inparaenum}
\item building a path for operator placement while optimizing one or more \ac{QoS} metrics (based on the knowledge from network discovery service),
\item placing the plan node with \ac{CEP} queries on the selected physical brokers and
\item collaboratively processing the deployed query and delivering the complex event.
\end{inparaenum}
In principle, this module can be extended to support different \ac{QoS} metrics, design characteristics and hence placement decisions.

\section{Evaluation} \label{sec:eval}

We evaluate the \system architecture by answering two questions: 

\textbf{EQ1} Is the \system system extensible and expressive?

\textbf{EQ2} How is the performance of \system system?

To this end, we explain the evaluation setup in Section \ref{subsec:setup}, \textbf{EQ1} in Section \ref{subsec:eq1} and \textbf{EQ2} in  Section \ref{subsec:eq3}.

\subsection{Evaluation Environment} \label{subsec:setup}
We selected the \ac{NFN} architecture~\cite{Tschudin2014} to implement our solution, due to its built-in support of resolving \emph{named functions} as so-called $\lambda$ expressions on top of the \ac{ICN} substrate. 
However, a major difference to our architecture is that the communication plane in \ac{NFN} is only consumer-initiated. 
In contrast, we provide unified communication layer for co-existing consumer- and producer-initiated interactions, while doing \ac{CEP} operations in the network. 
As a consequence, we embedded \ac{CEP} operators as \emph{named functions} while leveraging \ac{NFN}'s abstract machine to resolve them. NFN works together with CCN-lite~\cite{CCN-lite2019}, which is a lightweight implementation of CCNx and NDN protocol. We have developed unified interfaces of our design on top of NFN (v0.1.0) and CCN-lite (v0.3.0) for the Linux platform~\cite{INetCEPGithub2019}. 

We have enhanced the NDN protocol implementation in the CCN-lite and the NFN architecture by: 
\begin{inparaenum}
\item including the additional packet types and their handling, as described in \Cref{subsec:unified},
\item implementing the extensible general CEP query language, parser, and CEP operators as \ac{NFN} services, as described in Section \ref{subsec:language} and
\item implementing a network discovery service with modifications in both CCN and NFN, and operator placement as an NFN service (cf. Section \ref{subsec:ogprocessing}).
\end{inparaenum}

\begin{figure}
\centering
\subcaptionbox{Centralized processing.}
{\centering
\includegraphics[height=2.1cm, width=0.487\linewidth]{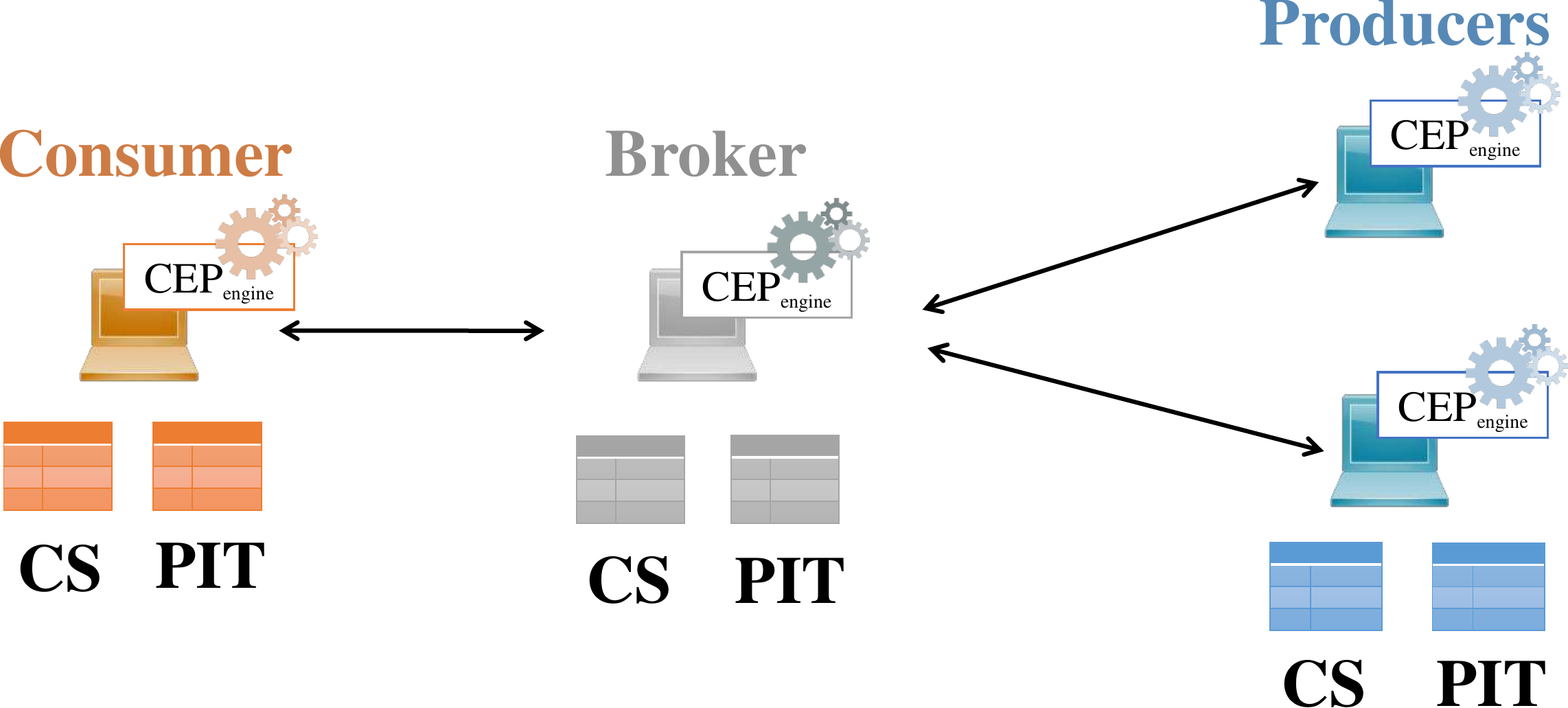}}
\hfill
\subcaptionbox{Distributed processing.}
{\centering
\includegraphics[height=3cm, width=0.49\linewidth]{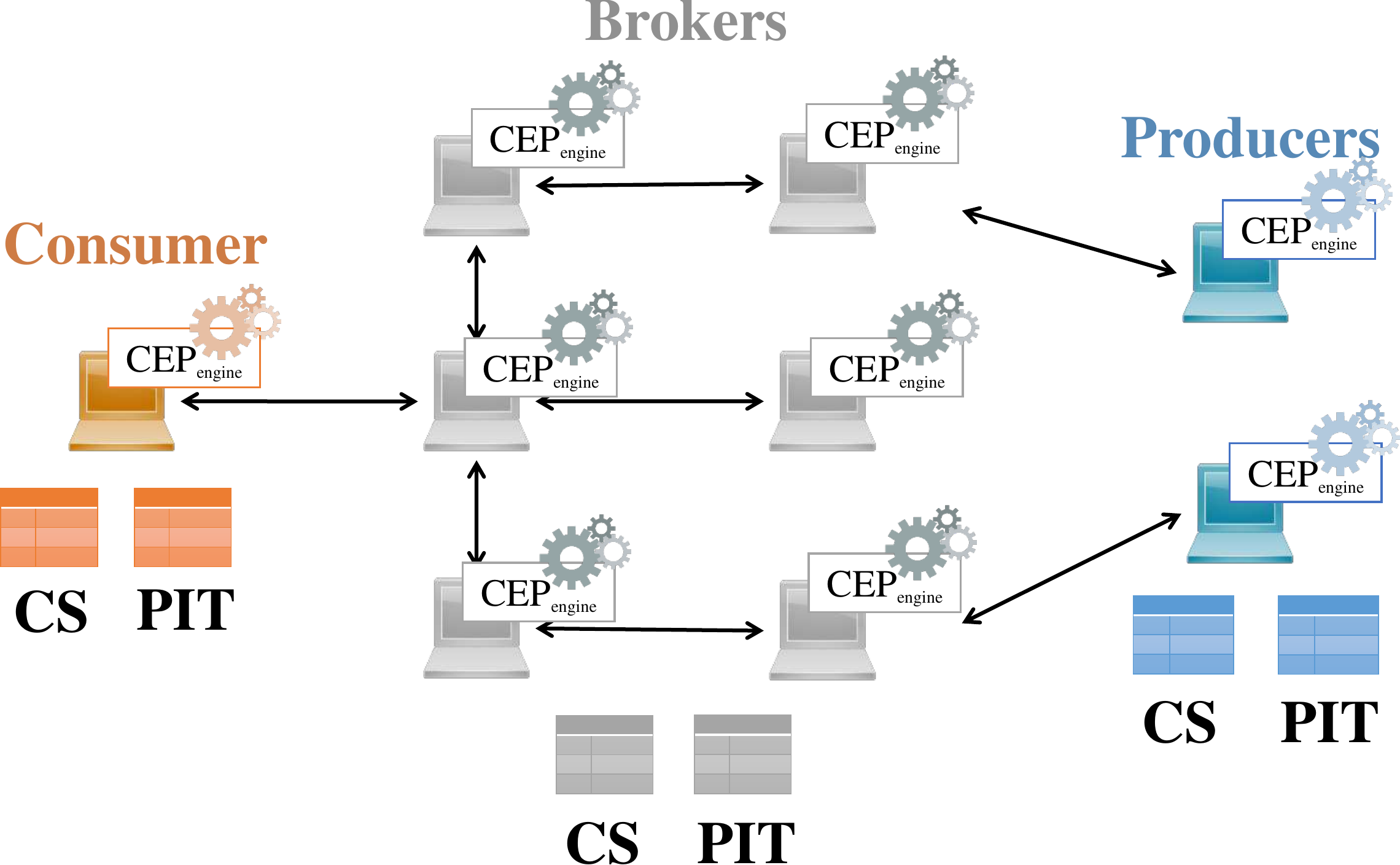}}
\setlength{\abovecaptionskip}{5pt}
\caption{Topology for evaluation.} \label{fig:topo}
\end{figure}

We evaluated our implementation using the CCN-lite emulator on two topologies: centralized (cf.~\Cref{fig:topo}a) and distributed (cf.~\Cref{fig:topo}b). Each node in our topology is an Ubuntu 16.04 virtual machine (VM) with 8 GiB of memory. Here, each VM (node) is a CCN-NFN relay, which hosts a \ac{NFN} compute server encapsulating the CEP operator logic. For running the experiments, we first created a CCN network topology as illustrated in~\Cref{fig:topo}. Second, we deployed the \system architecture that works on the NFN compute server, the CCN-NFN relay, as well as on the links. Here, as intended, the nodes communicate using the NDN protocol instead of IP. In the centralized topology, we have two producers, a single broker that processes the query and one consumer. In the distributed topology, we have one consumer, two producers and six brokers, as shown in the figure. 
The data structures \ac{CS} and \ac{PIT} are utilized as explained in the previous sections (cf.~\Cref{subsec:unified}). 
We use Queries~\ref{query:window}-\ref{query:join} (cf. \Cref{subsec:language}) for our evaluation with the DEBS grand challenge 2014 smart home dataset and the disaster field dataset. The dataset is explained in \Cref{subsec:eq1}.

\subsection{Evaluation Question I: Extensibility} \label{subsec:eq1}
To show the extensibility and expressiveness of our approach, we extended the \system architecture for the two representative \ac{IoT} use cases that we introduced in Section \ref{sec:motivation}, with a heat map query and a load prediction query.
We extended the \system query language and \ac{CEP} operators to include the heat map~\cite{Koepp2014} and prediction operators~\cite{Martin2014} by making a few additions to our implementation in our extensible query language and parser. 
We used real world datasets to evaluate the queries: the 2014 DEBS grand challenge and a disaster field dataset. 

\textbf{Dataset 1.} For the heat map query, we use a dataset~\cite{Alvarez2018} of a field test mimicking a post-disaster situation. The field test mimics two fictive events, a lightning strike and a hazardous substance release from a chemical plant, which resulted in a stressful situation. The collected dataset consists of sensor data, \eg location coordinates. It was collected from smartphones provided to the participants.
Each sensor data stream has a schema specifying the name of the attributes, \eg the GPS data stream has the following schema: 
\\
\begin{footnotesize}
$<ts, s\_id, latitude, longitude, altitude, accuracy, distance, speed>$
\end{footnotesize}

\begin{figure}
\centering
\includegraphics[height=3.2cm]{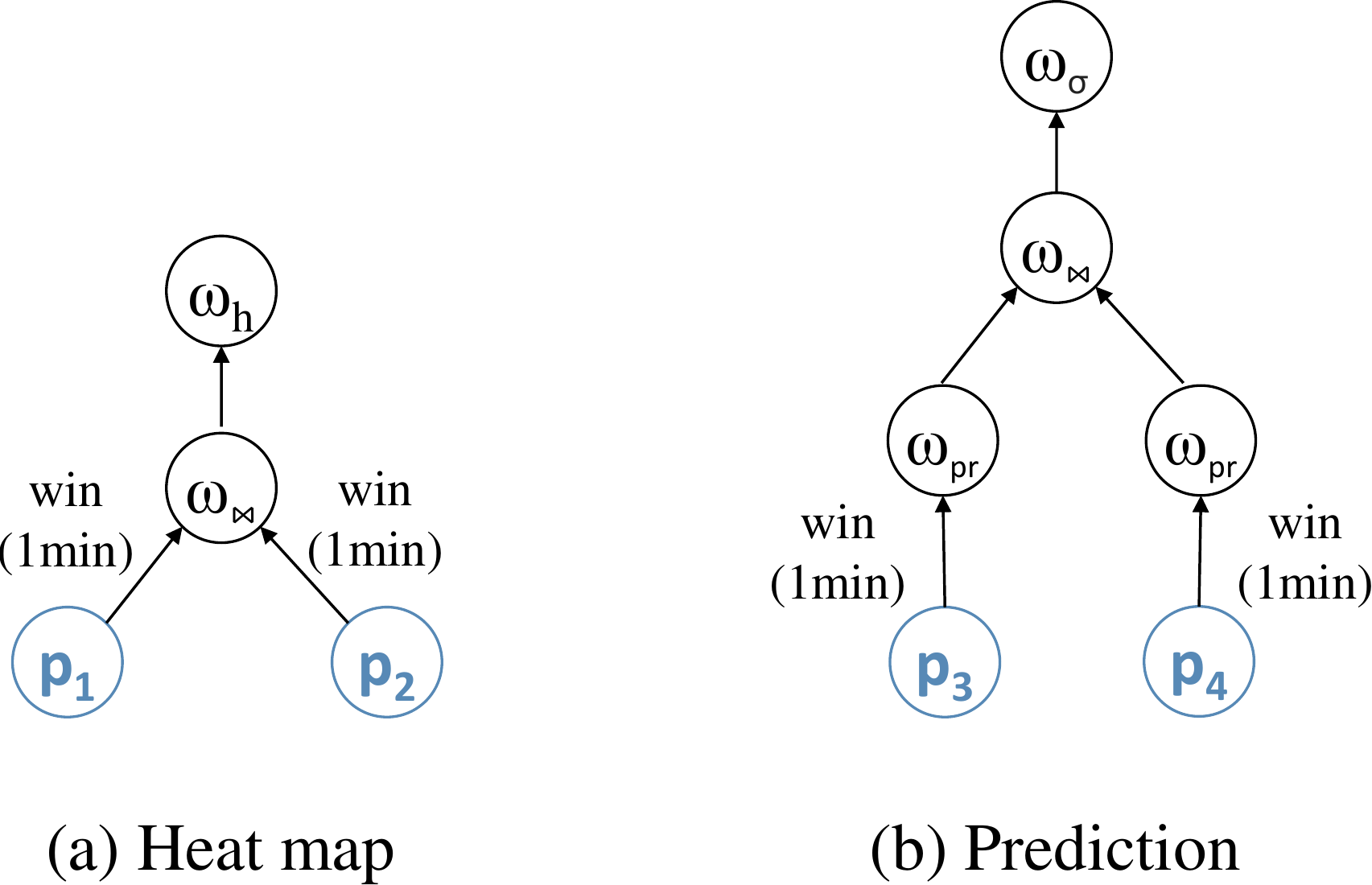}
\setlength{\abovecaptionskip}{5pt}
\caption{Two applications for evaluation (a) a heat map query for post-disaster relief and (b) an energy load forecasting query for smart homes.}
\label{fig:bothqueries}
\end{figure}

\textbf{Query.} We use the $latitude$ and $longitude$ attributes of this schema to generate the heat map distribution of the survivors from the disaster field test. A typical heat map application joins the GPS data stream from a given set of survivors, derives the area by finding minimum and maximum latitude and longitude values, and visualizes the heat map distribution of the location of the survivors in this area. For simplicity, we consider a data stream from two survivors, as shown in the operator graph in \Cref{fig:bothqueries}a. Here, $p_1$ and $p_2$ are the producers or GPS sensors, $\omega_{\Join}$ is the join operator, and $\omega_{h}$ is the heat map generation operator algorithm~\cite{Koepp2014}. This is easily possible using the \system language and parser implementation that follows an \emph{Abstract Factory} design pattern. First, we included the algorithm for heat map generation, which is $\tilde{20}$ LOC. Second, we extended the language implementation to include the user defined operator by adding $\tilde{20}$ LOC.

\begin{lstlisting}[caption={Display the heat map distribution of GPS source 1 and 2 in the given area with a given cell size.}, captionpos=b, label=query:heatmap]	
HEATMAP(
  'cell_size', 'area', 
  JOIN(WINDOW(GPS_S1, 1m), WINDOW(GPS_S2, 1m)) 
  GPS_S1.'ts' = GPS_S2.'ts')
)
\end{lstlisting}

\textbf{Dataset 2.} The second dataset comes from the 2014 DEBS grand challenge~\cite{DEBS2014} scenario focused on solving a short-term load forecasting problem in a smart grid. The data for the challenge is based on real-world profiles collected from smart home installations.
The dataset captured load measurements from 
unique smart plugs with the following schema: 
\\
\begin{footnotesize}
$<ts, id, value, property, plug\_id, household\_id, house\_id>$
\end{footnotesize}

\textbf{Query.} We apply an existing solution~\cite{Martin2014} to perform prediction by extending the \system architecture for two smart plugs.
The corresponding operator graph for such a prediction is illustrated in \Cref{fig:bothqueries}b and listed below. Here, $p_3$ and $p_4$ are the producers or smart plugs, $\omega_{\Join}$ is a join operator and $\omega_{pr}$ is a prediction operator based on the algorithm~\cite{Martin2014}. In the first query, we notify the consumer about the predictions for five minutes into the future, while in the second query we notify only if the predictions of load are above a threshold. Similarly to the heat map application, we implemented a prediction algorithm by adding $\tilde{50}$ LOC and the language implementation with $\tilde{20}$ LOC. 

The detailed description of the algorithms for the respective use cases in order to achieve the extensibility is presented in \Cref{apndix:extensibility}.
 
\begin{lstlisting}[caption={Performing a prediction every 5 minutes for 5 minutes into the future on the load observed by plug source 1 and 2.}, captionpos=b, label=query:predict]
JOIN(
  PREDICT(5m, WINDOW(PLUG_S1, 1m)),
  PREDICT(5m, WINDOW(PLUG_S2, 1m)) 
  PLUG_S1.'ts' = PLUG_S2.'ts'
)
\end{lstlisting}

\begin{lstlisting}[caption={Filter the load prediction values that are greater than 20 from Listing~\ref{query:predict}.}, captionpos=b, label=query:filterpredict]
FILTER(JOIN( 
  PREDICT(5m, WINDOW(PLUG_S1, 1m)), 
  PREDICT(5m, WINDOW(PLUG_S2, 1m)) 
  PLUG_S1.'ts' = PLUG_S2.'ts'
  ), 
  'load'>20)
\end{lstlisting}    		
 
\subsection{Evaluation Question II: Performance} \label{subsec:eq3}
We evaluated the performance of the \system architecture on standard \ac{CEP} queries including the Queries~\ref{query:window}-\ref{query:join}. Additionally, we evaluated the  heat map and prediction queries (Queries~\ref{query:heatmap}-\ref{query:filterpredict}).
Our aim is to understand the performance of: 
\begin{inparaenum}
\item the query parser on increasing the number of nested operators in operator graph, 
\item the operator graph module for different kind of queries, and
\item the operator placement module on increasing number of queries.
\end{inparaenum}

\begin{figure}[t]
\centering
\includegraphics[width=1.6in]{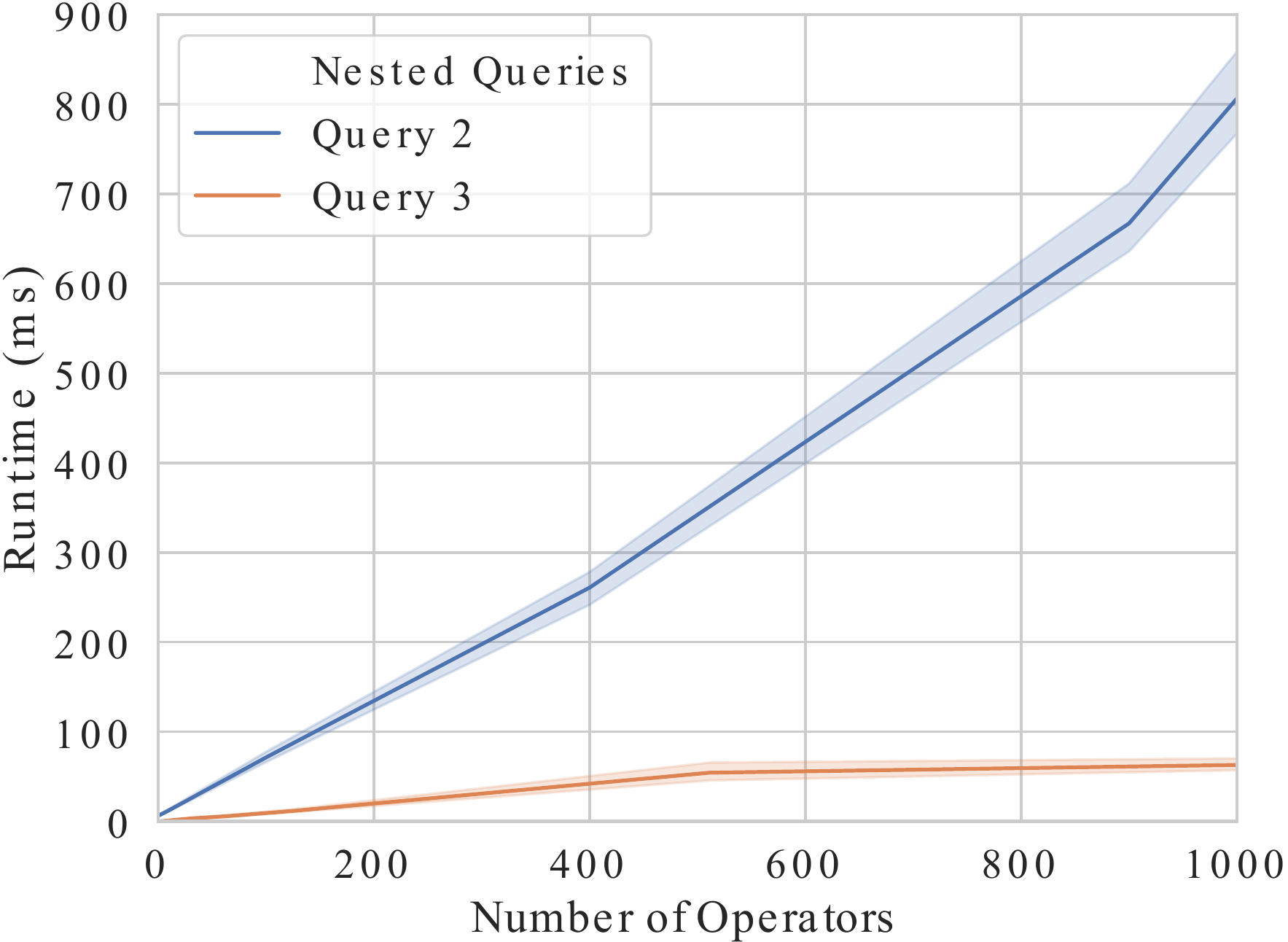}
\setlength{\abovecaptionskip}{3pt}
\caption{Performance of the query parser depending on increasing the number of nested operators.} \label{fig:parser}
\end{figure} 

\subsubsection{Query Parsing}
In the query parser design (cf. Section \ref{subsubsec:parser}), we performed a complexity analysis of the Algorithm~\ref{algo:parser}. In this section, we verify the analysis experimentally by increasing the number of operators in the operator graph while using the centralized topology shown in~\Cref{fig:topo}a.

\begin{figure}[t!]
\centering	
\subcaptionbox{End-to-end delay.\label{fig:q1toq6}}
{\includegraphics[width=1.5in]{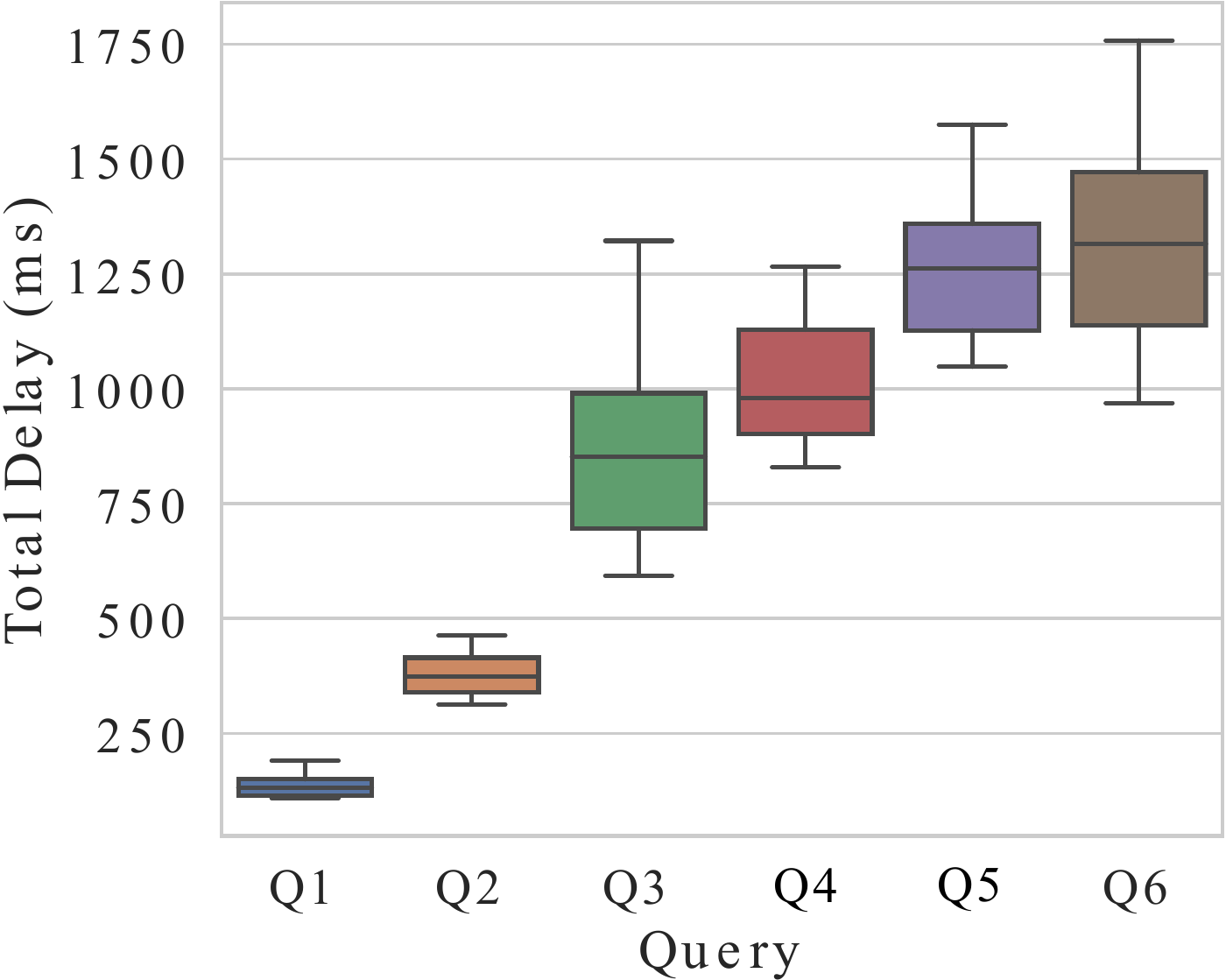}\vspace{-6pt}}
\hfill
\subcaptionbox{Communication delay.\label{fig:commq1toq6}}
{\includegraphics[width=1.5in]{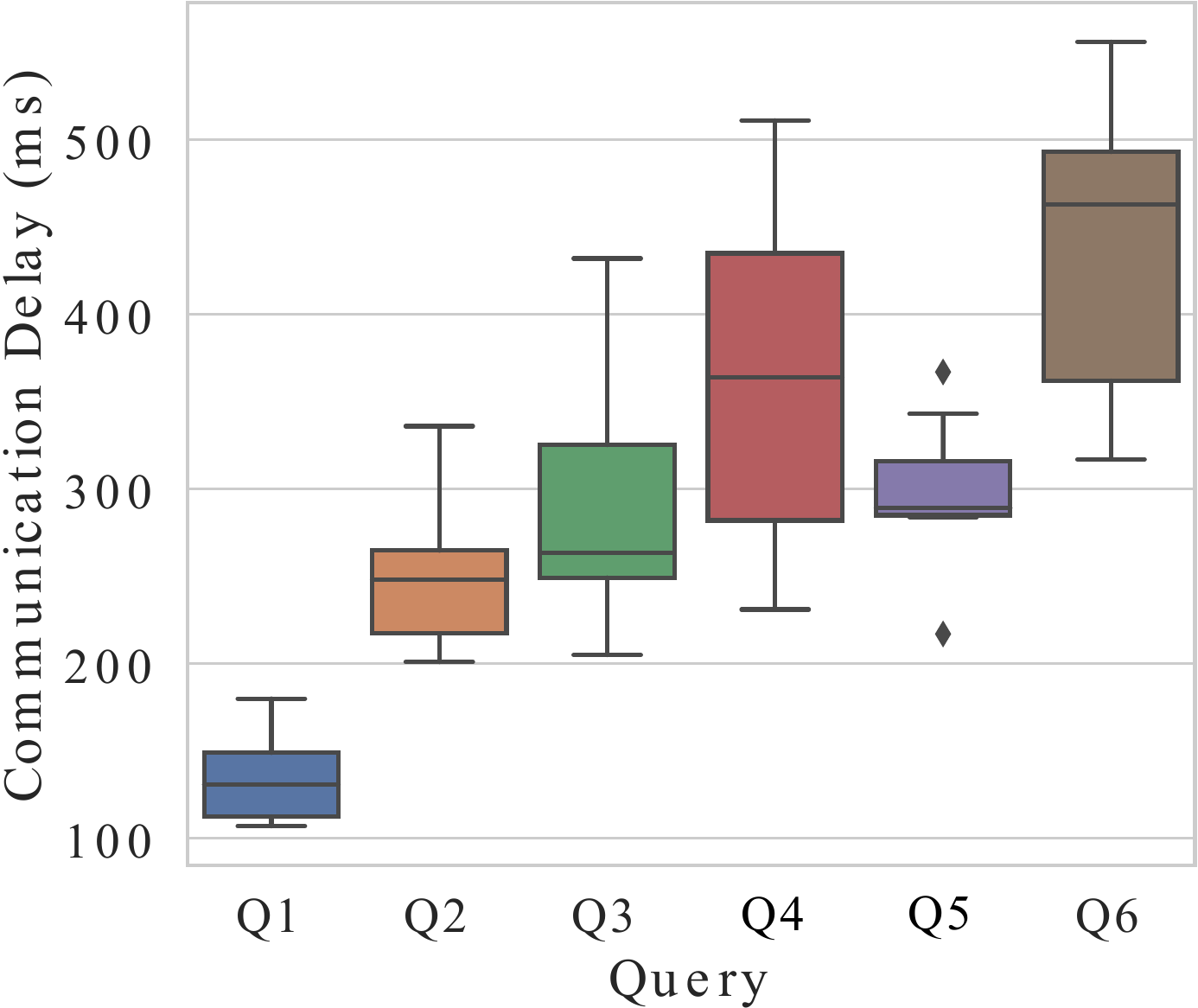}\vspace{-6pt}}
\setlength{\abovecaptionskip}{0pt}
\caption{End-to-end and communication delay observed in Queries 1 to 6.}
\label{fig:evaldelay}
\end{figure} 
In \Cref{fig:parser}, we show the performance of query parser algorithm in terms of runtime (in ms) for  Query~\ref{query:filter} or \texttt{Filter} operators and Query~\ref{query:join} or \texttt{Join} operators. We show in a line plot with a confidence interval of 95\% for 20 runs of the emulation that the two queries scales reasonably and can be processed in a few milliseconds.

\subsubsection{Centralized Query Processing}
We measured the end-to-end delay in processing the six queries defined above using the operator graph module (cf. \S \ref{subsec:ogprocessing}). In ~\Cref{fig:evaldelay}, we show the results as box plots with a confidence interval of 95\% for 10 runs of the emulation. For Query~\ref{query:window}-\ref{query:join}, the total delay perceived is less than a few milliseconds. It increased for the new queries Query~\ref{query:heatmap}-\ref{query:filterpredict}, where we introduced prediction and heat map operators, primarily due to increased consumption of data and the computational complexity of the algorithms for prediction~\cite{Martin2014} and heat map~\cite{Koepp2014}, respectively. We further show the distribution of the mean end-to-end delay in \Cref{tab:latency} to understand the primary reason for the delay. We listed the time spent in each of the modules of the \system architecture, namely, \emph{operator graph creation}, \emph{placement of operators}, and \emph{communication of events}. The values shown in the table are the mean of the values observed for 10 executions. For basic CEP queries ~\ref{query:window}-\ref{query:filter}, the major portion of time ($\tilde{98}\%$) is spent in communication.
This can be explained, as also confirmed in other works~\cite{Chen11}, by the limitations of the CCNx implementation of the NDN protocol. Hence, we observe the communication delays for all the queries in \Cref{fig:commq1toq6}, where it takes up to $500$ ms for delivering results of complex queries like prediction and heatmap.

\begin{table}
\centering
\footnotesize
\begin{tabular}{lcccc}
\hline
\textbf{Query} & \textbf{Total}   & \textbf{Graph}   & \textbf{Placement} & \textbf{Communication}   \\ \hline
\textbf{Query 1} &  132   & 1  & 0 & 131    \\
\textbf{Query 2} &  372.5  & 0.5 & 124 & 248  \\ 
\textbf{Query 3} &  850.5  & 1  & 586 & 263.5 \\ 
\textbf{Query 4} &  979.5  & 1 & 614.5 & 364  \\ 
\textbf{Query 5} &  1262  & 1 & 972 & 289 \\ 
\textbf{Query 6} &  1315  & 1  & 851 & 463  \\
\hline
\end{tabular}
\caption{The division of mean end-to-end delay in ms for of operator graph creation, placement of operators, and (communication) delay for centralized placement (see \Cref{fig:q1toq6}).} \label{tab:latency}
\end{table}

\begin{figure}[t]
\centering
\includegraphics[width=1.5in]{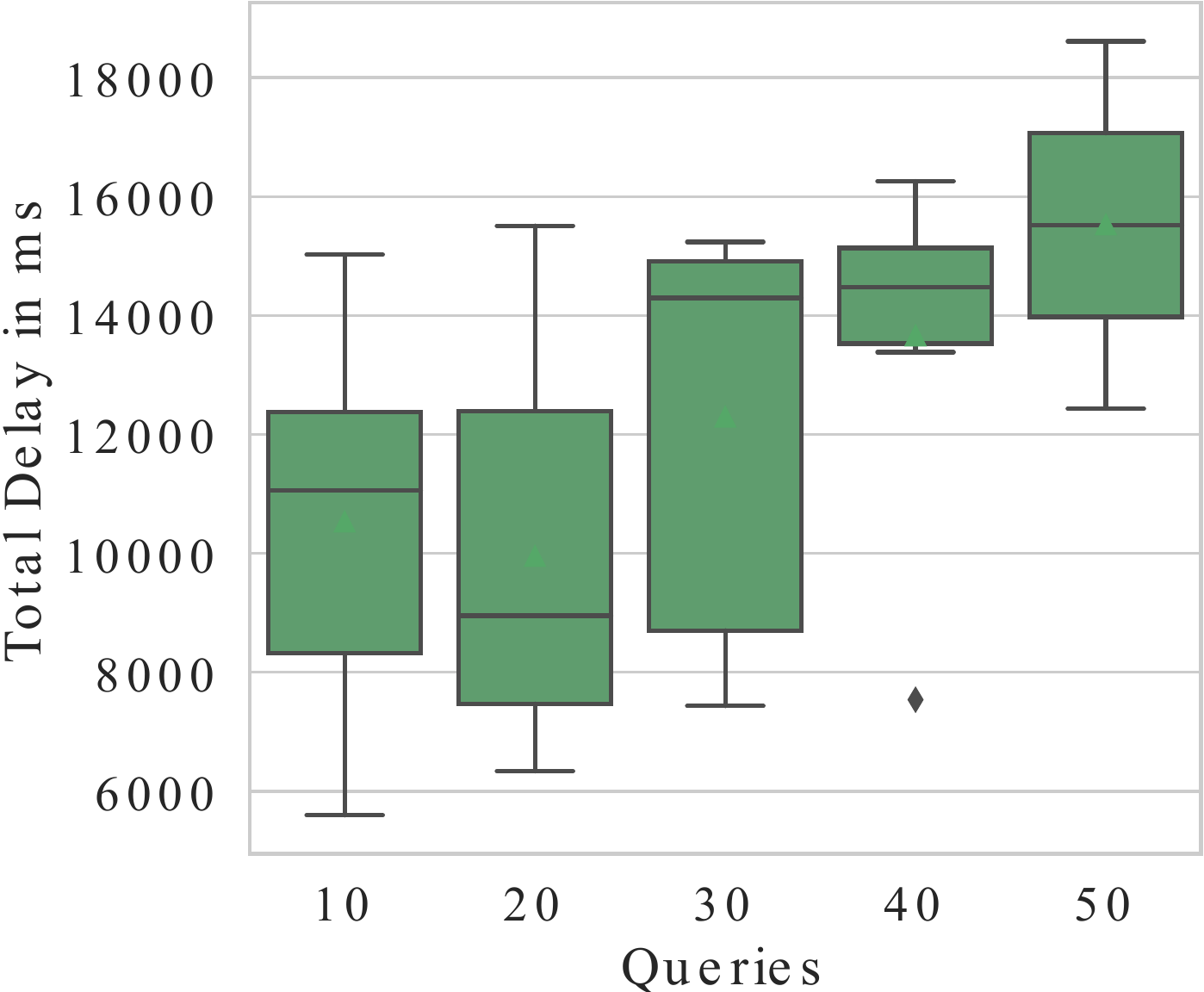}
\setlength{\abovecaptionskip}{0pt}
\caption{End-to-end delay observed for Query 3 on incrementally increasing the query workload.} \label{fig:multipleq3}
\end{figure} 

\subsubsection{Operator Placement}

To understand the behavior of the operator placement module, we utilize the distributed topology of seven VMs, as illustrated in \Cref{fig:topo}b, to place operators based on the information collected by the network discovery service. To take full advantage of distributed \ac{CEP}, we increased the query load of Query~\ref{query:join} starting from node 1 to 5. For example, the first 10 queries were initialized at the first broker node followed by the next 10 queries at the second node, and so on. Hence, the number of consumers also increased in the network with the query load. In~\Cref{fig:multipleq3}, we see that the total delay in retrieving the complex event increased incrementally with the query load,  which is reasonable given the static network size (7 nodes) and the resources. However, the queries were distributed evenly using the operator placement algorithm at distinct nodes.

To summarize, we evaluated the performance of \system on two topologies: centralized and distributed, using two \ac{IoT} applications and six different queries. Our evaluation shows that \ac{IoT} applications can be integrated seamlessly using the \system architecture, \ac{CEP} queries can be formulated and can be extended for more use cases, and simple \ac{CEP} queries can be processed in  milliseconds. 

\section{Related Work} \label{sec:relatedwork}
We now review the state-of-the-art \ac{ICN} architectures in terms of their support for consumer- and producer-initiated interaction patterns and existing \ac{INP} architectures in \Cref{subsec:icnarch}, and \ac{CEP} and networking architectures in \Cref{subsec:eparch}.

\subsection{ICN Architectures}~\label{subsec:icnarch}
\textbf{Interaction Patterns in ICNs:}
In Figure~\ref{fig:taxonomy}, we highlight the main \ac{ICN} architectures \ac{NDN}~\cite{Zhang2014a}, \ac{NFN}~\cite{Tschudin2014}, DONA~\cite{Koponen2007}, PURSUIT~\cite{PURSUITProject}, PSIRP~\cite{PSIRP}, in terms of their support of a unified communication layer as presented in our work. However, only those appearing in the green box, namely CONVERGENCE~\cite{Melazzi2010}, GreenICN~\cite{GreenICN}, and Carzaniga et al.~\cite{Carzaniga2011}, provide support for both kind of invocation mechanisms. The CONVERGENCE system combines the publish/subscribe interaction paradigm on top of an information-centric network layer. In contrast, we provide a unified interface such that pull and push based interaction patterns could co-exist in a single network layer while performing in-network computations. 
GreenICN is an \ac{ICN} architecture for post-disaster scenarios by combining NDN (pull-based) with COPSS~\cite{Chen11} (push-based). However, GreenICN introduces additional data structures, \eg  a subscription table (ST), while we provide this combination using the existing data structures of \ac{ICN}. Furthermore, it is not clear if GreenICN can function as a whole in a single \ac{ICN} architecture~\cite{GreenICN2016}. 
Carzaniga et al. propose a unified network interface similar to our work, however, the authors only propose a preliminary design of their approach~\cite{Carzaniga2011} without implementing it in an \ac{ICN} architecture, and subsequently focus on routing decisions~\cite{Carzaniga2013} rather than on distributed processing.

\textbf{INP and IoT Architectures in ICNs:}
Authors in work~\cite{Wang2016, Scherb2017} propose an approach to distribute computational tasks in the network by extending the \ac{NFN} architecture similar to our work. However, the authors do not deal with the lack of abstractions required for processing continuous data stream.
In contrast, we propose a unified communication layer to support \ac{CEP} over \ac{ICN} and an extensible query grammar and parser that opens a wide range of operators.
Krol et al.~\cite{Krol2017} propose NFaaS based on unikernels, which is a container based virtualization approach to encapsulate named functions placed on NDN nodes. However, they do not provide support for stateful functions, while \ac{IoT} functions can be stateful, e.g., involving time windows, which is supported by our architecture. 
Ahmed et al.~\cite{Ahmed2016} propose a smart home approach using \ac{NDN} and support both push and pull interaction patterns similar to our work. However, in their architecture they only support retrieving raw data, \eg humidity sensor readings, but not meaningful events as we do.
Shang et al.~\cite{Shang2016}
propose a publish/subscribe based approach for modern building management systems (BMS) in \ac{NDN}. However, the authors build on standard consumer-initiated interaction, as described in {Limitation 1} (cf.~\Cref{subsec:decisions}).
Publish-subscribe deployment for \ac{NDN} in the IoT scenarios has been discussed in previous works~\cite{Gundogan2018, ICNRG2017}.
These works confirm the need of integrating producer-initiated interaction in \ac{NDN}, however, do not provide a unified layer for both interaction patterns as we do.  

\subsection{\ac{CEP} and Networking Architectures}~\label{subsec:eparch}
\textbf{\ac{CEP} Architectures:} Several event processing architectures exist, ranging from, e.g., the open source Apache Flink~\cite{Carbone2015ApacheFS} to Twitter's Heron~\cite{Kulkarni2015} and Google's Millwheel~\cite{Akidau2013}.
One possibility is to interface one of them with an \ac{ICN} architecture. Initial work implemented Hadoop on \ac{NDN}~\cite{Gibbens2017} for datacenter applications. However, this requires changing the network model to push in contrast to our work, which would limit the support for a wide range of applications, as discussed above.

\textbf{Networking Architectures:} Another emerging network architecture is Software-Defined Networking (SDN)~\cite{Kreutz2015}, which is gradually being deployed, \eg in Google's data centers. It allows network managers to program the control plane to support efficient traffic monitoring and engineering. The {SDN} architecture is complementary to our work, since SDN empowers the control plane, while \ac{ICN} upgrades the data plane of the current Internet architecture.

\textbf{Data Plane and Query Languages:} The literature discusses many \ac{CEP} query languages~\cite{Carbone2015ApacheFS}. 
The novelty of the proposed query language is to allow for a mapping of operations to \ac{ICN}'s data plane. 
Alternative designs builds on P4~\cite{Bosshart2014} in the context of SDN. 
Initial work on programming \ac{ICN} with P4~\cite{Signorello2016} faced several difficulties due to lack of key language features and the strong coupling of the language to SDN's data plane model. 

\section{Discussion}\label{sec:discussion}
In this section, we discuss important future challenges that could be interesting to provide more sophisticated networking, reliability and optimization mechanisms in the \system architecture. 

\textbf{Sophisticated Flow and Congestion Control:} In \ac{CCN}, the \ac{PIT} table ensures the flow balance since one \Data packet is sent for each \Interest packet. The \DataStream packets of \system could disturb the rule of flow balance since the producer 
could overflow the buffer on the broker side. For this, \system implements a simple flow control mechanism where we restrict the receiver (consumer/broker) to specify maximum outstanding messages at a time. However, since the forwarding logic of \DataStream packets is similar to IP multicast, existing sophisticated multicast congestion control solutions like TCP-Friendly Multicast Congestion Control~\cite{IETF2006} and similar can provide sophisticated flow and congestion control.

\textbf{Reliability:} The brokers or consumers could miss packets when the available bandwidth and resources at their end is lower than the sending rate. This makes the presence of a module of reliability relevant, which can be catered by extending our work with existing reliable \ac{CEP} solutions~\cite{Koldehofe2013} or by looking into equivalent IP solutions such as Scalable Reliable Multicast~\cite{Floyd1997}.

\textbf{Query Optimization:} In \system we provide a placement module that maps the operator graph to in-network elements of \ac{CCN}. Another complementary direction could be to generate an optimal operator graph, \eg based on operator \emph{selectivity} or even partition operator graph by performing query optimization~\cite{Cao2013}. 

\textbf{Optimizing QoS:} In \system we provide a programming abstraction for the developers to write \ac{CEP} queries over \ac{ICN} data plane substrate. In addition, the placement module can be extended to look into further decentralized solutions and even other \ac{QoS} metrics~\cite{Cardellini2016} like throughput, availability, etc.

\section{Conclusion}\label{sec:conclusion}
In this paper, we proposed the \system architecture that implements a unified communication layer for co-existing consumer-initiated and producer-initiated interaction patterns. 
We studied important design challenges to come up with our design of a unified communication layer. In the unified layer, both \emph{consumer-} and \emph{producer-}initiated interaction patterns can co-exist in a single \ac{ICN} architecture. 
In this way, a wide range of \ac{IoT} applications are supported. 
With the proposed query language, we can express interest in aggregated data that is resolved and processed in a distributed manner in the network.
In our evaluation, we demonstrated in the context of two \ac{IoT} case studies that our approach is highly extensible.
The performance evaluation showed that queries are efficiently parsed and deployed, which yields - thanks to the in-network deployment - a low end-to-end delay, \eg simple queries induce only few milliseconds of overall delay.

Interesting research directions for future work are: (i) enhancing the performance of query processing by using parallelization, (ii) porting CCNx implementation on real hardware to accomplish low communication delays, and (iii) developing CEP compliant caching strategies. 

\section*{Acknowledgements}
This work has been co-funded by the German Research Foundation (DFG) as part of the project C2, A3 and C5 within the Collaborative Research Center (CRC) SFB 1053 -- MAKI.

\bibliographystyle{abbrv}
\bibliography{INetCEP-main-IEEE}

\appendix
\section{Appendix}
In this section, we define our query grammar in~\Cref{apndix:grammar}, the implementation details of our extensible design in~\Cref{apndix:extensibility} and the algorithms used for extending  the system with load forecasting and a heat map application in \Cref{apndix:applications}.

\subsection{Query Grammar}~\label{apndix:grammar}

\begin{definition}\label{def:grammar} A grammar consists of four components:
\begin{enumerate}
\item A set of \textbf{terminals} or \emph{tokens}. Terminals are the symbols that occur in a language.
\item A set of \textbf{non-terminals} or \emph{syntactic variables}. Each of them represent a set of strings. We define them the way we want to use them.
\item A set of \textbf{production} rules that define which non-terminals can be replaced by which terminals, or non-terminals or a combination of both. Here,  the terminal is the \textit{head} of the \textit{left side} of the production, and the replacement is the \textit{body} or \textit{right side} of the production. For example,
\textit{head} $\rightarrow$ \textit{body}
\item One of the non-terminals is designated as \textbf{start symbol} for each production.
\end{enumerate}
\end{definition}

Following the above definition~\ref{def:grammar} and Chomsky-Hierarchy~\cite{Hopcroft2006}, we selected a type 2 grammar or a \emph{context-free} grammar for \textsc{CEP over ICN} language since a query may consist of multiple subqueries (or operators), which can be expressed (out of many possible ways) using parenthesis "$()$".

\begin{table}
    \centering
    \footnotesize
    \begin{tabular}[c]{r l}
        \hline
        $\omega$  &  ::= $ \Join |  \sigma  |  win  |  agg  |  seq  $ \\
        $\Join$ &  ::=  \texttt{JOIN(  format  ,  $\omega$  ,  $\omega$  ,  boolExp  )}\\
        $\sigma$ &  ::=   \texttt{FILTER( format  ,  $\omega$  )}\\
        win &  ::=   \texttt{WINDOW( latinNumber  ,  number  )}\\
        seq & ::= \texttt{SEQUENCE( format  , $\omega \rightarrow$  $\omega$)} \\
        agg & ::= \texttt{AGGFUN(format, latinNumber, win)}  \\
        AGGFUN & ::= \texttt{SUM | MIN | MAX | AVG | COUNT} \\
        number  & ::=  \textit{REG([0-9]+)}  \\
        latinNumber  & ::=  \textit{REG([a-zA-Z0-9]+)} \\
        boolExp  & ::=  latinNumber  comparison  latinNumber  |  boolExp  concat  boolExp \\
        comparison &  ::=  $<  |  >  |  =  |  <=  |  >=$ \\
        concat &  ::=  \&  |  "|"  \\
        time & ::=  \textit{nn  :  nn  :  nn  .  nnn} \\
        n &  ::=  \textit{REG([0-9]\{1\})} \\
        format & ::=  \DataStream  |  \Data  \\
        \hline
    \end{tabular}
    \captionof{table}{Initial context-free grammar for \system language.}
    \label{tab:grammar}
\end{table}
The context-free grammar allows us to combine the production rules. The head of each production consist only of one non-terminal and the bodies are not limited by only one terminal and/or one non-terminal. This is because the language needs to embed operators in parenthesis and with a regular grammar we cannot have an arbitrary number of  parentheses. We need a way to memorize each parenthesis and this ability is given by context-free grammars.

We define an initial \system language grammar and represent it using BNF (Backus-Naur form) in \Cref{tab:grammar}, considering the aforementioned design decisions. We use regular expressions represented as $\mi{REG} (\ldots)$, where $(\ldots)$ can be literals $\mi{[a-z]}$ and $\mi{[A-Z]}$ in lower and upper case, respectively, and numbers $\mi{[0-9]}$.   
The plus ($+$) sign in $\mi{REG ([a-z]+)}$ means that at least one lowercase letter has to appear, while a $1$ in $\mi{REG ([a-z]\{1\})}$ means that exactly one lowercase letter has to appear. The relational operators given by $\mi{comparison}$ define a binary relation between two entities, \eg two column names of a schema, a column name to a number or a column index to a number.

\subsection{Extensibility}~\label{apndix:extensibility}
\begin{figure}[h]
    \centering
    \includegraphics[width=0.5\textwidth]{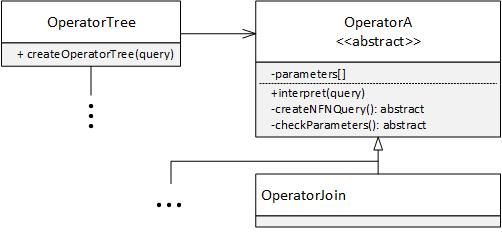}
    \caption{Operator Definition in UML}
    \label{fig:parserUml}
\end{figure}

To make our query language extensible, we follow a well-known \textit{Abstract Factory} design pattern from object-oriented programming for our operator definition as illustrated in~\Cref{fig:parserUml}. Algorithm~\ref{algo:parser} is the starting point of our operator graph creation. This is implemented in the \texttt{OperatorTree} class. Each operator inherits the abstract class \texttt{OperatorA} which defines the \texttt{interpret} (parseQuery in Algorithm) function as seen in the figure. The \texttt{checkParameters} verifies the correctness of the parameters.

If a new operator is to be included, it will override the existing methods of the abstract class and the parameters correctness has to be defined. This allows minimal changes in the implementation for each new application developed using our system.

\subsection{Applications}~\label{apndix:applications}

\subsubsection{Short term Load Forecasting}

For the DEBS Grand Challenge in 2014, Martin et al. derived requirements for predicting future energy consumption of a plug \cite{Martin2014}. We formulate our requirements with respect to our \system system:
\begin{enumerate}
    \item To meet the goal of making an estimation on future energy consumption, it is necessary to use historical data as a reference.
    \item In order to run on a machine with limited resources, the prediction algorithm needs to be lightweight in computation power and storage.
\end{enumerate}

\begin{figure}[t]
    \centering
    \includegraphics[width=0.4\textwidth]{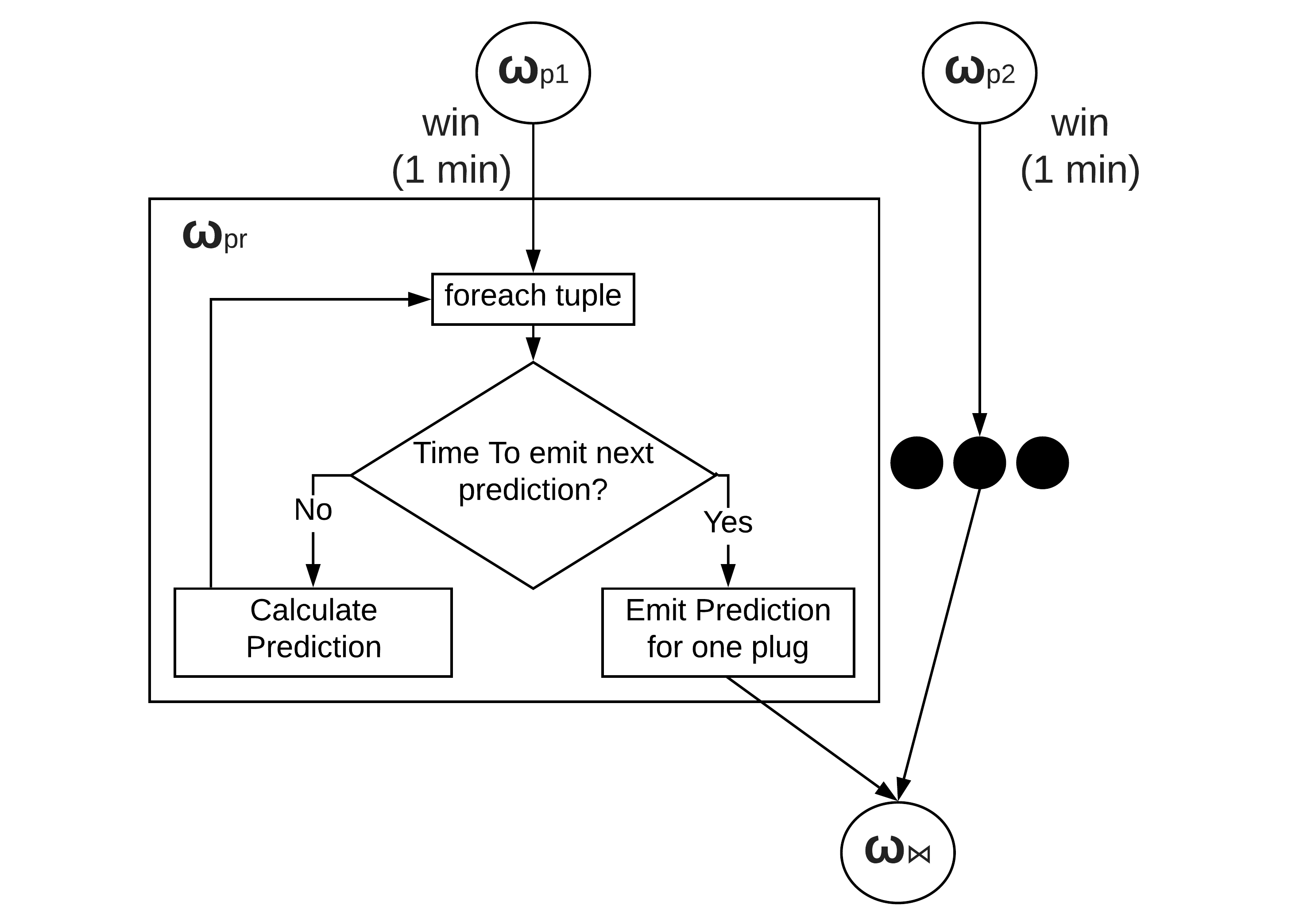}
    \caption{Flow chart explaining the prediction algorithm}
    \label{fig:prediction1_operator_flow}
\end{figure}

The formula for predicting future load is given by the publishers of the DEBS Grand Challenge and is as follows: \( predicted\_load\left( s_{i+2} \right) = \left( avgLoad\left( s_{ i } \right) + median\left( \left\{ avgLoads\left( s_{ j } \right) \right\} \right) \right) \), Here, \(i\) is the current timestamp, \(s_{i}\) the currently recorded values at time \(i\) and \(s_{j}\) the past values at a corresponding time \(j=i+2\). The load \emph{two steps} in the future is therefore made up of the current average electricity consumption and the average electricity consumption from the past.

In~\Cref{fig:prediction1_operator_flow}, we represent the flow of the prediction algorithm as per the requirements defined above and Query~\ref{query:predict}. For each time window of 1 minute, we first determine if it is the time for next prediction, which is provided as an input in the query. If the time has not come yet, a value for prediction (average load) is calculated and stored so that it can be used for the equation defined above. Inversely, if it is the time to make a prediction, a prediction tuple of the following form is emitted.

$<ts, plug\_id; household\_id; house\_id; predicted\_load>$  \\
Here, $ts$ is the timestamp of the prediction, $plug\_id$ identifies a socket in a household, $household\_id$ identifies a household within a house and $house\_id$ identifies a house and $predicted\_load$ is the prediction as specified in the equation above.

\subsubsection{Heat Map}
 Algorithm \ref{algo:heatmapAlgo} describes the heat map creation and visualization for the location updates from survivors of the disaster field test used in this work based on \cite{Koepp2014}.
\begin{algorithm}[h]
\small
\KwData{$loc$: Window of location $<lat, long>$ tuple of the survivor}
\KwData{$Lat_{min}$: The minimum latitude value}
\KwData{$Lat_{max}$: The maximum latitude value}
\KwData{$Long_{min}$: The minimum longitude value}
\KwData{$Long_{min}$: The maximum longitude value}
\KwData{$HC$: The number of horizontal cells needed to map the values}
\KwData{$VC$: The number of vertical cells needed to map the values}
\KwData{$cell\_size$: The granularity}
\KwData{$Grid$: A two dimensional array}
$HC$ = \(\lfloor{\frac{Long_{max}- Long_{min}}{cell\_size}}\rfloor \) \;
$VC$ = \(\lfloor{\frac{Lat_{max}- Lat_{min}}{cell\_size}}\rfloor \) \;
\ForEach{line in $S_D$}{
absLatVal = $loc$[lat] - $Lat_{min}$\;
absLongVal = $loc$[long] - $Long_{min}$\;
$Grid$[\( \lfloor{\frac{absLatVal}{cell\_size}}\rfloor\)][\( \lfloor{\frac{absLongVal}{cell\_size}}\rfloor\)] += 1\;
}
 \KwRet{$Grid$}
\newline
\caption{Algorithm for the heat map operator}
\label{algo:heatmapAlgo}
\end{algorithm}

In line 1, we calculate the number of horizontal cells required for the desired heat map. For this we divide the difference between the maximum and minimum longitude by the desired cell\_size, which indicates how large and finely meshed the resulting heat map should be and then round this value down to the next smaller number (given by floor function). In line 2, similarly we compute the vertical cells. For each of these location tuples in the current window, the first absolute latitude and longitude values are computed in line 4 and 5, respectively. By dividing these values by the cell\_size, we obtain the corresponding position in the heat map in line 6.

\end{document}